\PassOptionsToPackage{pdfpagelabels=false}{hyperref} 

\documentclass[nolinenumbers,twocolumn]{aastex63}

\usepackage{acronym}
\usepackage{hyperref}
\usepackage{amsmath,amsfonts,amssymb}
\usepackage{mathrsfs}
\usepackage{mathtools}
\usepackage[utf8]{inputenc}
\usepackage{CJK}
\usepackage[normalem]{ulem}
\usepackage{graphicx}
\graphicspath{{./}{../}}
\usepackage{pifont}

\newcommand{\gt}{\widetilde{\gamma}}
\newcommand{\At}{\widetilde{A}}
\newcommand{\Gt}{\widetilde{\Gamma}}
\newcommand{\Kh}{\widehat{K}}

\newcommand{\dd}{\partial}
\newcommand{\third}{\tfrac{1}{3}}

\newcommand{\Uvec}{\mathcal{U}}
\newcommand{\Pvec}{\mathcal{P}}
\newcommand{\Fvec}{\mathcal{F}}

\date{}

\shorttitle{{\sc Chana}: A GPU-Accelerated AMR GRRMHD Code}

\begin{document}
\title{{\sc Chana}: A GPU-Accelerated Adaptive-Mesh Code for General Relativistic Magnetohydrodynamics with Multigroup M1 Neutrino Transport} 

\correspondingauthor{Tianshu Wang}
\author[0000-0002-0042-9873]{Tianshu Wang}
\affiliation{Department of Physics, University of California, Berkeley, CA, 94720-7300 USA}
\email{tianshuw@berkeley.edu}

\begin{abstract}
This paper presents {\sc Chana}, a GPU-accelerated, block-structured adaptive-mesh code for general relativistic magnetohydrodynamics, multigroup M1 neutrino transport, and dynamical spacetime evolution using the Z4c formulation. Each physical module uses independently selected spatial discretizations and ghost-zone widths to balance accuracy and computational cost. Stiff radiation-matter interactions are treated implicitly, with Jacobians calculated analytically from the chain-rules. The code is written in C++20 and it uses MPI and Kokkos 5 for distributed-memory and performance-portable parallelism. The code is validated through a number of GRMHD, radiation transport, and dynamic spacetime tests. For a representative core-collapse supernova configuration with three neutrino species and 12 energy groups per species, the code achieves approximately $1.1\times10^6$ cell updates per GPU second on NVIDIA A100 GPUs using four-stage RK4 integration. Weak-scaling efficiency remains above 85\% up to 256 GPUs on Perlmutter.
\end{abstract}

\keywords{relativity -- magnetohydrodynamics (MHD) -- radiative transfer -- methods: numerical}

\section{Introduction}
\label{sec:intro}

Core-collapse supernovae (CCSNe) and binary neutron-star (BNS) mergers are among the most
energetic transient phenomena in the universe. In such systems, the coupled evolution of multiple physical processes jointly determines the final outcomes. 
General relativity (GR) governs the spacetime in which all other physical processes happen. Hydrodynamic instabilities such as convection and turbulence can aid the explosions and change the evolution of both compact objects and ejecta. Strong magnetic fields either inherited from progenitors or amplified by dynamos redistribute energy and angular momentum and may launch relativistic outflows. Neutrinos carry a major fraction of released energy, while neutrino-matter interactions exchange energy with the fluid and determine its electron fraction, with direct consequences on the dynamics and nucleosynthesis of the ejecta. Predictive numerical simulations of such systems must therefore combine
GR, magnetohydrodynamics (MHD), and radiation transport in a single framework.
In addition, these systems naturally span a large range of spatial scales. The compact central objects, shocks, and (magneto)hydrodynamic instability structures all require high spatial resolution, while the computational domain must extend far enough to follow the expanding ejecta. Mesh refinement concentrates resolution where it is needed and is therefore essential for extending multi-physics calculations to spatial scales and evolution times relevant for astrophysical phenomena. Adaptive refinement allows the mesh to follow evolving structures as compact objects move, shocks propagate, and ejecta expand. The computational cost of these coupled calculations also motivates implementations that exploit GPU-based supercomputers.

Significant efforts have been made along such paths. General-relativistic
MHD codes have reached high levels of robustness and parallel scalability and some 
have developed mature GPU-accelerated AMR implementations \citep{Liska2022,Stone2020,Kalinani2025,stone2026,Han2026,Musolino2026}.
Multi-group radiation-hydrodynamics codes with approximate GR treatments, frequently equipped with static mesh refinement (SMR), have been well established and proven successful in CCSN simulations, yet only a small number of them have been ported to GPUs
\citep{muller2010,Just2015,skinner2019,Bruenn2020,Melson2020,Endeve2026}.
When combined with fully relativistic dynamical spacetime, more approximate radiation transport schemes such as grey M1 and multi-group M0 methods are commonly used \citep{Foucart2015,radice2022,barker2024,Palenzuela2025}. A smaller number of CPU codes that evolve general-relativistic radiation-hydrodynamics with multi-group M1 radiation transport have also been developed in the past years \citep{Roberts2016,kuroda2016,Anninos2020}, but the high computational cost of the scheme remains a bottleneck. Very recently, \citet{Kuroda2026} presented ANRe-M1, a GPU-accelerated code coupling dynamical spacetime evolution to multi-group M1 neutrino radiation hydrodynamics on a fixed mesh-refinement hierarchy, which is a performance-portable upgrade of the \citet{kuroda2016} code.

\begin{CJK*}{UTF8}{gbsn}
This paper presents {\sc Chana} 
(刹那, ``in an instant''), a GPU-accelerated adaptive mesh refinement
code that evolves general relativistic magnetohydrodynamics, multi-group M1 neutrino transport, and dynamical
spacetime using the Z4c formulation within a coupled numerical framework. The physics modules may
be evolved jointly, or separately with the remaining sectors disabled or held fixed. The code allows each physical module 
to select its own spatial order and ghost widths, allowing computational cost and accuracy to be balanced separately for each module.
The code is written in C++20. MPI provides distributed-memory parallelism, while Kokkos 5 \citep{Trott2022}
provides on-node execution on CPUs and GPUs. When available, GPU-aware MPI can be utilized.
\end{CJK*}

The paper is arranged as follows.  Section~\ref{sec:eq} gives the physical equations solved by the code.
Section~\ref{sec:num} describes the numerical implementation, and
Section~\ref{sec:tests} presents the verification tests and performance
measurements. Finally, Section~\ref{sec:summary} summarizes the results.

\section{Equations}
\label{sec:eq}

We summarize here the physical equations for the coupled system.

\subsection{Conventions and units}
\label{sec:conventions}

The equations below are written in geometrized units, $G=c=1$, with metric
signature $(-,+,+,+)$. Greek indices run over spacetime and Latin indices run over
space.  

The spacetime is decomposed as
\begin{equation}
  ds^2=-\alpha^2dt^2+\gamma_{ij}(dx^i+\beta^i dt)(dx^j+\beta^jdt)
\label{eq:adm}
\end{equation}
with $n^\mu=\alpha^{-1}(1,-\beta^i)$ being the normal observer. Here $\alpha$, $\beta^i$, and $\gamma_{ij}$ are the lapse, shift, and spatial
metric, respectively.  We write $\gamma=\det\gamma_{ij}$, and denote by $D_i$ the covariant derivative compatible with $\gamma_{ij}$, so that for a spatial
vector $D_iV^i=\gamma^{-1/2}\dd_i(\gamma^{1/2}V^i)$.  The extrinsic curvature of the slice is 
\begin{align}
  &K_{ij}
        =-\frac{1}{2\alpha}\left(\dd_t\gamma_{ij}
             -\beta^k\dd_k\gamma_{ij}
             -2\gamma_{k(i}\dd_{j)}\beta^k\right),
  \nonumber\\
  &K=\gamma^{ij}K_{ij}.
  \label{eq:extrinsic}
\end{align}

\subsection{General-relativistic magnetohydrodynamics}
\label{sec:grmhd}

The fluid is described by its fluid-frame density $\rho$, pressure $p$, specific internal energy $\epsilon$, electron fraction $Y_e$, spatial coordinate components $u^i$ of the fluid four-velocity, and magnetic four-vector measured in the fluid frame $b^\mu$. The total fluid-magnetic stress-energy tensor and the dual electromagnetic field tensor are then written as:
\begin{align}
  &T_{\rm MHD}^{\mu\nu}
   =(\rho h+b^2)u^\mu u^\nu
    +\left(p+\frac{b^2}{2}\right)g^{\mu\nu}-b^\mu b^\nu, 
  \nonumber\\
        &(^\ast F)^{\mu\nu} = b^\mu u^\nu-b^\nu u^\mu.
  \label{eq:mhd-F}
\end{align}
The coupled GRMHD equations are:
\begin{align}
  &\nabla_\mu(\rho u^\mu)=0,
       \nonumber\\
  &\nabla_\mu T_{\rm MHD}^{\mu\nu}
       =\sum_{s}\int d\varepsilon G_{s,\varepsilon}^{\nu},
       \nonumber\\
  &\nabla_\mu(\rho Y_e u^\mu)
       =-\sum_{s}\int d\varepsilon x_{s,\varepsilon}\,u_\mu G_{s,\varepsilon}^{\mu},
       \nonumber\\
  &\nabla_\mu(^\ast F)^{\nu\mu}=0
       \label{eq:induction}
\end{align}
where $^\ast F$ is the dual electromagnetic tensor, $G_{s,\varepsilon}^{\mu}$ is the radiation four-force on the matter, and $x_{s,\varepsilon}$ converts the exchanged neutrino energy into electron lepton number. The sign of $x_{s,\varepsilon}$ is positive for $\nu_e$, negative for $\bar\nu_e$, and zero for heavy-lepton neutrinos. Accurate lepton-number conservation without separately evolving neutrino number requires a compatible discretization, which is described in Section~\ref{sec:num-m1}. The subscript $s$ runs over the different radiation species and $\varepsilon$ is the fluid-frame photon/neutrino energy. The equation of state closes the fluid system through $p=p(\rho,\epsilon,Y_e)$.

Let
\begin{equation}
  v^i=\frac{u^i}{W}+\frac{\beta^i}{\alpha},\,\,
  W=\left(1-\gamma_{ij}v^iv^j\right)^{-1/2},\,\,
  \bar v^i=\alpha v^i-\beta^i,
  \label{eq:eulerian-velocity}
\end{equation}
be the fluid three-velocity measured by the Eulerian observer, its Lorentz
factor, and the coordinate transport velocity, and let
$B^i=-({}^{\ast}F)^{i\mu}n_\mu$ be the magnetic field measured by that observer (the factor $4\pi$ is absorbed into the field's normalization). The above GRMHD equations can be put into a more useful form:
\begin{align}
  &\dd_t\left(\sqrt{\gamma}D\right)
    +\dd_j\left(\sqrt{\gamma}D\bar v^j\right)=0,\nonumber\\
  &\dd_t\left(\sqrt{\gamma}DY_e\right)
    +\dd_j\left(\sqrt{\gamma}DY_e\bar v^j\right)
    =-\alpha\sqrt{\gamma}\sum_{s} \int d\varepsilon x_{s,\varepsilon}\,u_\mu G_{s,\varepsilon}^{\mu},\nonumber\\
  &\dd_t\left(\sqrt{\gamma}S_i\right)
    +\dd_j\left[\sqrt{\gamma}\left(\alpha S_i{}^j-\beta^jS_i\right)\right]
    \nonumber\\ &=\sqrt{\gamma}\left[\frac{\alpha}{2}S^{jk}\dd_i\gamma_{jk}
        +S_k\dd_i\beta^k
        -\left(\tau+D\right)\dd_i\alpha\right]\nonumber\\
        &\quad+\alpha\sqrt{\gamma}\sum_{s}\int d\varepsilon\gamma_{i\mu}G_{s,\varepsilon}^{\mu},\nonumber\\
  &\dd_t\left(\sqrt{\gamma}\tau\right)
    +\dd_j\left\{\sqrt{\gamma}\left[
        \alpha\left(S^j-Dv^j\right)-\beta^j\tau\right]\right\}\nonumber\\
    &=\alpha\sqrt{\gamma}\left[S^{jk}K_{jk}
        -S^j\dd_j\ln\alpha\right]-\alpha\sqrt{\gamma}\sum_{s}\int d\varepsilon n_\mu G_{s,\varepsilon}^{\mu},\nonumber\\
  &\dd_t(\sqrt{\gamma}B^i)
+\partial_j\!\left[\sqrt{\gamma}
(B^i\bar v^j-B^j\bar v^i)\right]=0,\nonumber\\
  &\partial_i(\sqrt\gamma B^i)=0.
    \label{eq:grmhd-eqs}
\end{align}
where 
\begin{align}
  &D=\rho W, \nonumber\\
  &S_i=\left(\rho hW^2+B^2\right)v_i-\left(B^kv_k\right)B_i,
  \nonumber\\
  &S_{ij}=\left(\rho hW^2+B^2\right)v_iv_j
       +\left(p+\frac{b^2}{2}\right)\gamma_{ij}
       -\frac{B_iB_j}{W^2}\nonumber\\
       &\,\,\,\quad-\left(B^kv_k\right)\left(B_iv_j+B_jv_i\right),\nonumber\\
  &h=1+\epsilon+p/\rho\nonumber\\
  &\tau=\rho hW^2-p+B^2-\frac{b^2}{2}-D.
  \label{eq:grmhd-cons}
\end{align}

\subsection{General-relativistic M1 neutrino transport}
\label{sec:eq-m1}

For each species $s$ and fluid-frame energy $\varepsilon=-u_\mu k^\mu$ where $k^\mu$ is the photon/neutrino four-momentum, the radiation stress-energy tensor
is decomposed with respect to the Eulerian observer as
\begin{align}
  &R_{s,\varepsilon}^{\mu\nu}=E_{s,\varepsilon}n^\mu n^\nu
       +n^\mu F_{s,\varepsilon}^{\nu}+n^\nu F_{s,\varepsilon}^{\mu}
       +P_{s,\varepsilon}^{\mu\nu},\nonumber\\
  &n_\mu F_{s,\varepsilon}^{\mu}=n_\mu P_{s,\varepsilon}^{\mu\nu}=0,
  \label{eq:radiation-decomposition}
\end{align}
where $E$, $F^\mu$, and $P^{\mu\nu}$ are the Eulerian radiation energy density,
flux, and pressure tensor per unit energy. The zeroth and first angular moments obey
\begin{equation}
  \nabla_\nu R_{s,\varepsilon}^{\mu\nu}
  -\frac{\dd}{\dd\varepsilon}
     \left(\varepsilon Q_{s,\varepsilon}^{\mu\nu\lambda}
                         \nabla_\lambda u_\nu\right)
  =-G_{s,\varepsilon}^{\mu}.
  \label{eq:radiation-moments}
\end{equation}
in which the energy derivative involves the third-rank moment $Q_{s,\varepsilon}^{\mu\nu\lambda}$. 

Writing the equations in a similar conservative form as Eq.~\eqref{eq:grmhd-eqs} (suppressing the $s$ and $\varepsilon$ subscripts), one gets
\begin{align}
  &\dd_t\left(\sqrt{\gamma}E\right)
   +\dd_j\left[\sqrt{\gamma}\left(\alpha F^j-\beta^jE\right)\right]
   +\alpha\sqrt{\gamma}\,
     \dd_\varepsilon\left(\varepsilon\,n_\mu Q^\mu\right)\nonumber\\
   &={}\alpha\sqrt{\gamma}\left[P^{jk}K_{jk}
        -F^j\dd_j\ln\alpha\right]+\alpha\sqrt{\gamma}\,n_\mu G^{\mu},
   \nonumber\\
  &\dd_t\left(\sqrt{\gamma}F_i\right)
   +\dd_j\left[\sqrt{\gamma}\left(\alpha P^j{}_i-\beta^jF_i\right)\right]
   -\alpha\sqrt{\gamma}\,
     \dd_\varepsilon\left(\varepsilon\,\gamma_{i\mu} Q^\mu \right)\nonumber\\
   &={}\sqrt{\gamma}\left[\frac{\alpha}{2}P^{jk}\dd_i\gamma_{jk}
        +F_k\dd_i\beta^k-E\,\dd_i\alpha\right]-\alpha\sqrt{\gamma}\,\gamma_{i\mu}G^{\mu},
   \nonumber\\
   &Q^\mu= Q^{\mu\nu\lambda}\nabla_\lambda u_\nu.
\end{align}

\subsection{Dynamical spacetime}
\label{sec:spacetime}
The code evolves Einstein's equations in the Z4c formulation of \citet{Bernuzzi2010}. The implementation
closely follows the equations and gauge choices in \citet{Daszuta2021} and \citet{zhu2025}, which are briefly listed below:
 
Write $R_{ij}$ for the Ricci
tensor of $\gamma_{ij}$ and $R=\gamma^{ij}R_{ij}$ for its trace, and define the
matter projections
\begin{align}
  &\mathcal{E}=n_\mu n_\nu T^{\mu\nu},\,\,
  \mathcal{S}_i=-\gamma_{i\mu}n_\nu T^{\mu\nu},\nonumber\\
  &\mathcal{S}_{ij}=\gamma_{i\mu}\gamma_{j\nu}T^{\mu\nu},\,\,
  \mathcal{S}=\gamma^{ij}\mathcal{S}_{ij},
  \label{eq:matter-projections}
\end{align}
in which the energy-momentum-stress tensor $T^{\mu\nu}$ contains both GRMHD and radiation contributions, and 
introduce
\begin{equation}
  \begin{aligned}
    &\chi = \gamma^{-1/3},\,\
    \gt_{ij}=\chi\gamma_{ij},\,\,\nonumber\\
    &\At_{ij}=\chi\left(K_{ij}-\third\gamma_{ij}K\right),\,\,
    \Kh=K-2\Theta,\nonumber\\
    &\Theta = -n_\mu Z^\mu,\,\, \Gt^i = 2\gt^{ij}Z_{j}+\gt^{ij}\gt^{kl}\dd_l\gt_{jk}
  \end{aligned}
  \label{eq:Z4c-vars}
\end{equation}
with $\det\gt_{ij}=1$, $\gt^{ij}\At_{ij}=0$ and $Z_j$ the spatial components of the auxiliary dynamical vector field in the Z4 formulation,
the Z4c equations are then written as
\begin{align}
  &\dd_t\chi={}
       \frac{2}{3}\chi\left(\alpha K-\dd_k\beta^k\right)+\beta^k\dd_k\chi,
       \nonumber\\
  &\dd_t\gt_{ij}={}-2\alpha\At_{ij}
       +\beta^k\dd_k\gt_{ij}
       -\frac{2}{3}\gt_{ij}\dd_k\beta^k
       +2\gt_{k(i}\dd_{j)}\beta^k,
       \nonumber\\
  &\dd_t\Kh={}-D^iD_i\alpha
       +\alpha\left(\At_{ij}\At^{ij}+\third K^2\right)
       +\beta^k\dd_k\Kh\nonumber\\
     &\,\quad\quad\quad+\alpha\kappa_1(1-\kappa_2)\Theta
       +4\pi\alpha(\mathcal{E}+\mathcal{S}),
       \nonumber\\
  &\dd_t\At_{ij}={}
       \chi\left[-D_iD_j\alpha+\alpha R_{ij}
                    -8\pi\alpha\mathcal{S}_{ij}\right]^{\rm TF}\nonumber\\
     &\,\,\quad\quad\quad+\alpha\left(K\At_{ij}-2\At_{ik}\At^k{}_j\right)
       \nonumber\\
     &\,\,\quad\quad\quad+\beta^k\dd_k\At_{ij}+2\At_{k(i}\dd_{j)}\beta^k
       -\frac{2}{3}\At_{ij}\dd_k\beta^k,
       \nonumber\\
  &\dd_t\Gt^i={}2\alpha\left[
       \widetilde{\Gamma}^{i}_{jk}\At^{jk}
       -\frac{3}{2\chi}\At^{ij}\dd_j\chi-\third\gt^{ij}\dd_j(2\Kh+\Theta)
       -8\pi\gt^{ij}\mathcal{S}_j\right]\nonumber\\
       &\quad\quad\quad-2\At^{ij}\dd_j\alpha
       -2\alpha\kappa_1
       \left(\Gt^i-\gt^{jk}\widetilde{\Gamma}^{i}_{jk}\right)\nonumber\\
       &\quad\quad\quad+\beta^k\dd_k\Gt^i-\gt^{mn}\widetilde{\Gamma}^{k}_{mn}\dd_k\beta^i\nonumber\\
       &\quad\quad\quad+\frac{2}{3}\gt^{mn}\widetilde{\Gamma}^{i}_{mn}\dd_k\beta^k
       +\third\gt^{ij}\dd_j\dd_k\beta^k
       +\gt^{jk}\dd_j\dd_k\beta^i,
       \nonumber\\
  &\dd_t\Theta={}\beta^k\dd_k\Theta\nonumber\\
       &\,\,\,\,\quad\quad+\alpha\left[\frac{1}{2}\left(R+\frac{2}{3}K^2
                         -\At_{ij}\At^{ij}-16\pi\mathcal{E}\right)-(2+\kappa_2)\kappa_1\Theta\right],
       \label{eq:Z4c-Theta}
\end{align}
where TF denotes the trace-free part with respect to the conformal metric. The following
$1+\log$ lapse variant of the puncture gauge is used:
\begin{align}
  \dd_t\alpha &= \beta^k\dd_k\alpha-2\alpha\Kh,
      \nonumber\\
  \dd_t\beta^i &= \beta^k\dd_k\beta^i+\Gt^i-\eta_\beta\beta^i.
       \label{eq:gauge-shift}
\end{align}

The intrinsic curvature $R_{ij}$ is decomposed as $\widetilde{R}^\chi{}_{ij}+\widetilde{R}_{ij}$, where
\begin{align}
    &\widetilde{R}^\chi{}_{ij} = \frac{1}{2\chi}\left(\widetilde{D}_i\widetilde{D}_j\chi+\gt_{ij}\widetilde{D}^l\widetilde{D}_l\chi-\frac{1}{2\chi}(\widetilde{D}_i\chi)( \widetilde{D}_j\chi)\right)\nonumber\\
    &\quad\quad\quad\, - \frac{3}{4\chi^2}(\widetilde{D}^l\chi)(\widetilde{D}_l\chi)\gt_{ij},\nonumber\\
    &\widetilde{R}_{ij} =-\frac{1}{2}\gt^{lm}\dd_l\dd_m\gt_{ij}+\gt_{k(i}\dd_{j)}\Gt^k+\gt^{mn}\widetilde{\Gamma}^{k}_{mn}\Gt_{(ij)k}\nonumber\\
    &\quad\quad\,\,\,\, +\gt^{lm}(2\Gt^{k}_{l(i}\Gt_{j)km}+\Gt^k_{im}\Gt_{klj}),
\end{align}
with $\widetilde{D}_i$ compatible with $\gt_{jk}$.

\section{Numerical Implementation}
\label{sec:num}

\subsection{Mesh}
\label{sec:mesh}

The coordinate system in {\sc Chana} is logically Cartesian and is tiled by uniform mesh blocks. Each block contains $N_{\rm b}^3$ active cells, where the even block size $N_{\rm b}$ is selected at compile time.  Adaptive refinement uses an octree: a refined block is replaced 
by eight children with half the cell spacing in each direction, and neighboring 
leaf blocks are kept within one refinement level. Each block stores its face-, edge-, and corner-neighbor relations, including periodic images and coarse-fine interfaces. Therefore, a block may have up to 56 neighbors that it needs to communicate with. {\sc Chana} supports up to four independent refine/coarsen indicators at once. The currently implemented estimators include \citet{Lohner1987}-type normalized second-difference indicators on any selected cell-centered variables, indicators based on fractional change across cells of any selected cell-centered variables, an indicator based on pressure-scale-height, an indicator based on Mach number change, a Z4c $\dd\chi$ indicator, and a $\min \chi$ indicator.

Blocks are padded by ghost zones filled by direct exchange at equal resolution
and by restriction or prolongation across refinement boundaries, or by physical
boundary conditions at the edge of the domain. The size of ghost regions is chosen to be module-dependent
because different reconstruction methods can be used by different physical modules (MHD, radiation, and Z4c) at the same time, 
which will be discussed with more details next.

\subsection{Discretization, Reconstruction, Restriction, and Prolongation}
\label{sec:recon}

The code currently supports the following reconstruction methods for cell-centered variables: piecewise constant; piecewise linear, 
with either the minmod or the monotonized-central (MC) limiter \citep{Mignone2014}; and 
weighted essentially non-oscillatory (WENO) schemes including 3rd-order WENO3 \citep{Jiang1996}, 5th-order WENO-Z \citep{Borges2008}, 
and 7th-order WENO7 \citep{Gerolymos2009}.
The code allows different variables to use different discretization/reconstruction methods and ghost widths. This flexibility
allows the cheaper piecewise linear reconstruction methods to be used for the expensive radiation sector, 
while maintaining a higher spatial order for the Z4c sector. By default, the spatial differencing in the Z4c module uses sixth-order centered stencils on four ghost layers.

Prolongation and restriction are done in a similar flexible way. Following \citet{zhu2025}, the Z4c variables are prolongated and restricted by Lagrange polynomials over stencils that fully utilize the available ghost layers (order $N_g+1$ for $N_g$ ghost layers, so 5th order at coarse-fine interface when 6th order differencing is used). The restriction of hydrodynamic and radiation conservative variables uses volume averaging, while their prolongation is done using the piecewise linear method with the minmod limiter by default. This choice limits the spatial order of the radiation-magnetohydrodynamics to second order at coarse-fine interfaces, but is more stable as minmod creates unphysical structures around the interfaces much less frequently than sharper methods \citep{Stone2020}. The restriction of face-centered magnetic fields is done via a face-average, while the prolongation uses the method of \citet{toth2002}. The cell-centered magnetic fields of the prolongated/restricted cells are then derived from the face-centered variables.

\subsection{Time integration}
\label{sec:time}
The default integrator is the ten-stage, fourth-order explicit strong-stability-preserving (SSP) Runge-Kutta (RK) method eSSPRK$(10,4)$ of
\citet{Ketcheson2008}. Although this method needs more stages per step than the standard RK4, its larger stability region allows larger time step choices, which makes the actual cost per unit simulation time comparable while keeping the SSP property. In terms of memory cost, this method has a low-storage form which needs only two saved state vectors beyond the current solution vector. The standard RK4 integrator is also supported. The per-step computation cost of each time integrator is roughly proportional to the number of stages per step.

\subsection{Riemann solver}
\label{sec:riemann}

At each face, an orthonormal frame adapted to the face normal is constructed following \citet{Kiuchi2022}. The Riemann problem is solved in that frame and the resulting flux is transformed back and densitized. The code supports relativistic HLLE/HLLC/HLLD solvers. The relativistic HLLC/HLLD solvers are implemented closely following \citet{Kiuchi2022}. These advanced Riemann solvers involve a nonlinear root finding which may not converge, and if that happens the Riemann solver falls back to the more stable HLLE case. The HLLC/HLLD solvers can develop instabilities near strong grid-aligned shocks. In non-relativistic HLLD solvers, \citet{minoshima2021} stabilizes the solver by introducing a shock-detection factor, whose generalization to relativistic cases needs further study. Currently, whenever a shock is detected, the code switches to HLLE in a buffer region covering the shock. The size of the buffer region depends on the size of ghost halos.

\subsection{Constrained Transport}
\label{sec:emf}

The MHD discretization carries three staggered magnetic quantities: edge-centered EMF
${\mathcal E}$, face-centered primary magnetic field
$B_{\rm f}$, and cell-centered derived magnetic field
$B_{\rm c}$. At each stage of the time integrator, the face-centered magnetic field is updated by the EMFs on the four edges of the face, which are constructed with the upwind CT prescription of \citet{Gardiner2005}. This keeps the magnetic field divergence $\dd_i(\sqrt\gamma B^i)$ unchanged to round-off.  After the face field is advanced, $B_{\rm c}$ is derived from the adjacent face values and then coupled to the cell-centered fluid state. 
At a refinement boundary the coarse edge EMF value is overwritten by the average
of the fine sub-edges that lie on it. Similarly, the coarse face fluxes are overwritten by the area average of fine faces' fluxes. 

In addition to the ideal-MHD EMFs, the code supports resistivity and $\alpha$-dynamo terms. An arbitrary edge- or cell-centered function of the primitive state can be defined and added to the ideal-MHD EMFs. For cell-centered functions, the edge will get the average over its four incident cells.

\subsection{M1 Closure}
\label{sec:closure}
To close the M1 radiation transport equations, a closure relation that calculates $P^{\mu\nu}_{s,\varepsilon}$ and $Q^{\mu\nu\lambda}_{s,\varepsilon}$ from $E_{s,\varepsilon}$ and $F^\mu_{s,\varepsilon}$ is needed. This is typically done by interpolating between the optically thin and thick limits using the Eddington factor $\xi$ (e.g., \citet{Shibata2011}). Importantly, to reproduce the correct diffusion limit, this Eddington factor generally needs to be evaluated in the fluid frame, while the Eulerian to fluid frame transformation depends on the closure itself. This leads to a non-linear root-finding problem whenever a frame transformation is needed, which complicates the calculation especially when it's involved in the implicit radiation-matter coupling and results in a nested nonlinear root-finding problem.

This issue can be partly solved if the closure itself is Lorentz covariant and can be evaluated in the Eulerian frame to avoid the root finding. One example of Lorentz covariant closures is the Levermore closure \citep{Levermore1984}, because the assumption it's derived from,  that there exists a unique frame (the radiation rest frame) in which the radiation field is isotropic, is Lorentz covariant. At diffusion limit, the Levermore closure automatically gives the correct behavior because the fluid frame itself becomes the radiation rest frame, even if the closure is evaluated in the Eulerian frame. In addition to this physical argument, a numerical test of diffusion-limit on a moving medium is presented in Section~\ref{sec:moving_diffusion}. Because of its Lorentz covariance, the Levermore closure is the default choice in {\sc Chana}, while the standard nonlinear root-finding for a general closure choice is also implemented following \citet{radice2022}. Although there are a number of other closure choices in the literature \citep{murchikova2017}, \citet{2023ApJ...943...78W} found that closure choices produced comparatively small changes in the CCSN simulations. The importance of closure choices in other astrophysical settings remains uncertain.

When the Levermore closure is used, the pressure tensor is closed algebraically in the Eulerian frame, where flux factor and Eddington factor are calculated as
\begin{align}
    &f=\frac{\sqrt{\gamma_{ij}F^iF^j}}{cE},\,\,
    \xi=\frac{3+4f^2}{5+2\sqrt{4-3f^2}}
  \label{eq:flux-factor}
\end{align}
and the pressure tensor is expressed as 
\begin{equation}
  P^{\alpha\beta}=E\left[
       \frac{1-\xi}{2}\gamma^{\alpha\beta}
       +\frac{3\xi-1}{2}\frac{F^\alpha F^\beta}{\gamma_{ij} F^iF^j}\right].
  \label{eq:m1-closure}
\end{equation}

The third moment needed by the energy-space source is constructed from the
same stress tensor in the fluid frame \citep{Shibata2011}. The fluid-frame zeroth ($J$), first ($H^\mu$), and
second moments ($L^{\mu\nu}$) are written as:
\begin{align}
  &J=R^{\mu\nu}u_\mu u_\nu,\,\,
  H^\mu=-R^{\mu\nu}u_\nu-Ju^\mu,\nonumber \\
  &L^{\mu\nu}=R^{\mu\nu}-Ju^\mu u^\nu
       -H^\mu u^\nu-H^\nu u^\mu.
  \label{eq:fluid-frame-moments}
\end{align}
The fluid-frame flux factor $f_{\rm fl}$ used by the third-moment
closure is
\begin{equation}
  \begin{aligned}
    f_{\rm fl}&=\frac{\sqrt{h_{\mu\nu}H^\mu H^\nu}}{J},\,\,
    \xi_{\rm fl}=\frac{3+4f_{\rm fl}^2}{5+2\sqrt{4-3f_{\rm fl}^2}},\\
    C_{\rm thin}&=\frac{3\xi_{\rm fl}-1}{2},\,\,
    C_{\rm thick}=\frac{3(1-\xi_{\rm fl})}{2}.
  \end{aligned}
  \label{eq:fluid-frame-flux-factor}
\end{equation}
where $h^{\mu\nu}=g^{\mu\nu}+u^\mu u^\nu$ is the spatial projector
orthogonal to $u^\mu$.
Note that the expressions of $C_{\rm thin}$ and $C_{\rm thick}$ are not directly derived from the radiation-rest frame assumption of the Levermore closure 
and don't automatically give the correct fluid-frame diffusion-limit. This is why the flux factor has to be evaluated in the fluid frame.
In principle, these coefficients can be derived self-consistently from the Lorentz covariant Levermore closure \citep{2011JQSRT.112.1323V},
but for simplicity we choose this approximate expression.
The third angular moment is then
\begin{align}
  Q^{\alpha\beta\gamma}&=Ju^\alpha u^\beta u^\gamma + H^\alpha u^\beta u^\gamma
   +H^\beta u^\alpha u^\gamma +H^\gamma u^\alpha u^\beta \nonumber\\
   &+L^{\alpha\beta}u^\gamma+L^{\alpha\gamma}u^\beta+L^{\beta\gamma}u^\alpha\nonumber\\
   &+C_{\rm thin}J\frac{H^\alpha H^\beta H^\gamma}
      {(h_{\mu\nu}H^\mu H^\nu)^{3/2}}\nonumber\\
   &+\frac{C_{\rm thick}}{5}
      \left(h^{\alpha\beta}H^\gamma+h^{\alpha\gamma}H^\beta
           +H^\alpha h^{\beta\gamma}\right).
  \label{eq:m1-third-moment}
\end{align}

\subsection{Radiation fluxes}
\label{sec:num-m1}
The default method to calculate the spatial numerical flux at a face is the second-order method in \citet{radice2022}:
\begin{align}
  &\Fvec_{i+1/2}=\frac{\Fvec_i+\Fvec_{i+1}}{2}
   -A(1-\phi)c_{\rm LF}\frac{\Uvec_{i+1}-\Uvec_i}{2}
  \label{eq:m1-central-flux}
\end{align}
where $c_{\rm LF}$ coefficient is the larger of the light-cone speeds in the
two cells sharing the face,
\begin{equation}
  c_{\rm LF}=c\left(\alpha\sqrt{\gamma^{dd}}+|\beta^d|\right), 
  \label{eq:m1-light-cone}
\end{equation}
the optical-depth factor is
\begin{equation}
  A=\min\left(1,\frac{1}{\Delta x\,(\kappa + \sigma)}\right),
  \label{eq:m1-ap-factor}
\end{equation}
with $\kappa$ and $\sigma$ being the absorption and scattering opacities,
and the flux limiter is computed via
\begin{equation}
  \phi=\min\left[1,\max\left(0,
       \min\left(\frac{U_i-U_{i-1}}{U_{i+1}-U_i},
                  \frac{U_{i+2}-U_{i+1}}{U_{i+1}-U_i}\right)\right)\right].
  \label{eq:m1-limiter}
\end{equation}

When the above fluxes are used, the radiation module doesn't require any reconstruction onto cell faces and is cheaper and more diffusive than the HLL fluxes calculated from face states reconstructed by the PLM method which is also second order. Stronger diffusion is a double-edged sword: it makes the simulation significantly more stable, but it's also less accurate. Importantly, the flux formula of Eq.~\eqref{eq:m1-central-flux} is intrinsically second-order and cannot benefit from higher-order reconstruction methods. Therefore, the HLL fluxes are also implemented for completeness,  which can be coupled to an arbitrary order of reconstruction method. However, the fluxes of Eq.~\eqref{eq:m1-central-flux} are still the default choice because of their performance advantages.

The frequency-space drift is advanced separately as a conservative flux on the
internal energy-group edges, following the method of \citet{kuroda2016}.
With $\varepsilon_g$ the group center and $X_g$ the moment per unit energy (representing
the fluid-frame moments $X\in\{J,H^\mu,L^{\mu\nu}\}$), the face
value is
\begin{align}
  &X_{g+1/2}=X^L_g+X^R_{g+1},\nonumber\\
  &X^L_g=\frac{\Delta\varepsilon_g\,w_g}{\varepsilon_{g+1}-\varepsilon_g}X_g,
  \,\,
  X^R_g=\frac{\Delta\varepsilon_g\,(1-w_g)}{\varepsilon_g-\varepsilon_{g-1}}X_g,
  \label{eq:m1-frequency-face}
\end{align}
where the weights are
\begin{align}
  &w_g=\frac{f_{g+1/2}}{f_{g-1/2}+f_{g+1/2}},
  \nonumber\\
  &f_{g+1/2}=\left(\frac{E_g}{\Delta\varepsilon_g\varepsilon_g^3}\right)^{1-r}
            \left(\frac{E_{g+1}}{\Delta\varepsilon_{g+1}\varepsilon_{g+1}^3}\right)^{r},
  \label{eq:m1-frequency-weight}
\end{align}
with $r=(\varepsilon_{g+1/2}-\varepsilon_g)/(\varepsilon_{g+1}-\varepsilon_g)$ and $E_g=\int_{\varepsilon_{g-1/2}}^{\varepsilon_{g+1/2}}E_{\varepsilon}d\varepsilon$
the group-integrated energy density. 

The drift constructed from the moments at group faces is:
\begin{align}
  &D^\alpha_{g+1/2}=Q^{\alpha\beta\gamma}_{g+1/2} u_{\beta;\gamma},
  \nonumber\\
  &\mathcal{A}_{g+1/2}=n_\nu D^\nu_{g+1/2},
  \,\,
  \mathcal{A}_{i,g+1/2}=\gamma_{i\mu}D^\mu_{g+1/2},
  \label{eq:m1-frequency-drift}
\end{align}
Since $D^\alpha$ is linear in the moments at fixed $u_{\beta;\gamma}$ and the weights
are scalars shared by all moments, this is equivalent to applying the weights of
Eq.~\eqref{eq:m1-frequency-face} to $\mathcal{A}_g$ and $\mathcal{A}_{i,g}$
directly, which is how it is evaluated in the code.

The energy-space derivative is discretized as
$\partial_\varepsilon(\varepsilon X)\to
[\varepsilon_g(X_{g+1/2}-X_{g-1/2})+\Delta\varepsilon_g X_g]/\Delta\varepsilon_g$.  Substituting
Eq.~\eqref{eq:m1-frequency-face} turns it into a difference of edge fluxes in
which each side is weighted by the center energy of the opposite group,
\begin{align}
  &\Phi^E_{g+1/2}=\varepsilon_{g+1}\mathcal{A}^L_g+\varepsilon_g\mathcal{A}^R_{g+1},
  \nonumber\\
  &\Phi^{F_i}_{g+1/2}=\varepsilon_{g+1}\mathcal{A}^L_{i,g}+\varepsilon_g\mathcal{A}^R_{i,g+1}.
  \label{eq:m1-frequency-flux}
\end{align}
These fluxes are then multiplied by $\alpha\sqrt{\gamma}\,\Delta t$
when applied to the conserved radiation state.  The above implementation maintains energy, 
momentum, and number conservation to good accuracy \citep{kuroda2016}.

\subsection{Radiation-matter coupling}
\label{sec:coupling}

The interaction terms are expressed using the fluid-frame moments $J$ and $H^\mu$, whose calculation can be found in Equation \eqref{eq:fluid-frame-moments}. For each species and energy group, the four-force is
\begin{equation}
  G^\mu_{s,\varepsilon}=\left(\kappa_{s,\varepsilon} J_{s,\varepsilon}-\eta_{s,\varepsilon}\right)u^\mu
       +\left(\kappa_{s,\varepsilon}+\sigma_{s,\varepsilon}\right)H^\mu_{s,\varepsilon},
  \label{eq:four-force}
\end{equation}
where $\eta$, $\kappa$, and $\sigma$ are the emissivity, absorption opacity, and scattering opacity supplied by the opacity interface. 
Such four-force terms are solved implicitly in each RK stage. The unknowns $\Pvec$ are the six fluid variables
$(\rho,v^i,T,Y_e)$ together with Eulerian-frame $(E,F_i)$ for every species and group, so the
system has $6+4N_sN_g$ components. Note that the set of unknowns $\Pvec$ is different from either the primitive or conservative variables. 
This is to reduce the amount of work in the implicit solve, since the EOS and opacity tables in the code are usually tabulated in $(\rho, T, Y_e)$ space. The convergence criterion is based on the maximum relative residual in the Eulerian frame ($10^{-6}$ by default). 

Although the Jacobian can be computed from numerical differencing, it is not the preferred method because the expensive residual function will be called a lot of times per iteration.
Whenever possible, the code tries to (semi-)analytically calculate the Jacobian. This benefits from the fact that all the table
interpolations (EOS, opacity, and when applicable, the closure) are made to return not only the variable values but also the derivatives. As a result, the Jacobian is derived using the chain-rule and additional residual evaluations required by finite differencing are avoided.

In most cases, the Jacobian matrix has a particular structure: each group's $4\times4$ $(E,F_i)$ block is independent of
every other group, and groups communicate only through the six fluid columns. The solve therefore factors each $4\times4$ block, eliminates it, and reduces
to a $6\times6$ Schur complement in the fluid variables, at a cost linear in the number of groups rather than cubic. But this faster solver will no longer work if energy groups are coupled together via processes like inelastic scattering. 

Before doing the full Newton-Raphson solve, cheaper candidate steps using approximate residual functions are attempted first if the cell meets certain criteria. Candidates are accepted only when they satisfy the same exact-residual tolerance as the full nonlinear solve. Two candidate estimators have been implemented. The first one focuses on the low-velocity limit, and the second assumes the Eddington factor is unchanged during the radiation-matter interaction processes. These candidate steps cost little extra time compared to the exact solver, but they will save a significant amount of time if candidates are accepted. Therefore, the actual performance depends on the acceptance rates of such candidates, and strongly coupled regions in general cost more than optically-thin regions. To avoid imbalanced workload distribution, a timer module which measures the actual computational cost of each MPI rank is implemented, whose results are fed to the workload balance module.

\subsection{Optimization}

The performance-critical loops in {\sc Chana} are written with Kokkos
\citep{Trott2022}, with the local physics routines marked as device-callable so
that the same source is compiled for host and accelerator back ends.  MPI
distributes mesh blocks among ranks, while Kokkos parallelizes the blocks owned
by each rank.  Volume kernels use a single flattened iteration space over
blocks and cells (and, where appropriate, variable components), rather than a
separate launch for each block.  This exposes enough parallelism even when
$N_{\rm b}$ is small. 

Variables use a component-major structure-of-arrays layout, with the cells of
one component contiguous in memory.  Neighboring threads therefore access
neighboring $k$ cells of the same variable, giving coalesced accesses on GPUs. 
Several bandwidth-bound operations are fused: in particular, reconstruction, the Riemann solve, and
flux divergence can be evaluated in one cell kernel, and each Runge--Kutta
update combines all required stage vectors with the right-hand side in one
pass.  Temporary device buffers are retained and reused across stages, avoiding
allocations and their associated synchronization inside the step loop.

The packing and unpacking of each block's ghost halo are also Kokkos kernels.  A team handles one halo
segment and vectorizes its contiguous inner direction. Same-rank traffic avoids
an additional copy between send and receive buffers, while inter-rank transfers
use device buffers with GPU-aware MPI when available and pinned-host staging
otherwise.  Time-step
constraints, error checks, and similar global diagnostics are formed with
device reductions, so only their small scalar results are returned to the
host.

\section{Test Problems}
\label{sec:tests}
This section presents several standard tests of the {\sc Chana} code, covering the relativistic MHD,
the radiation transport, and the dynamic spacetime modules.

\subsection{Linear Wave Tests}

\begin{figure}
  \centering
  \includegraphics[width=0.96\columnwidth]{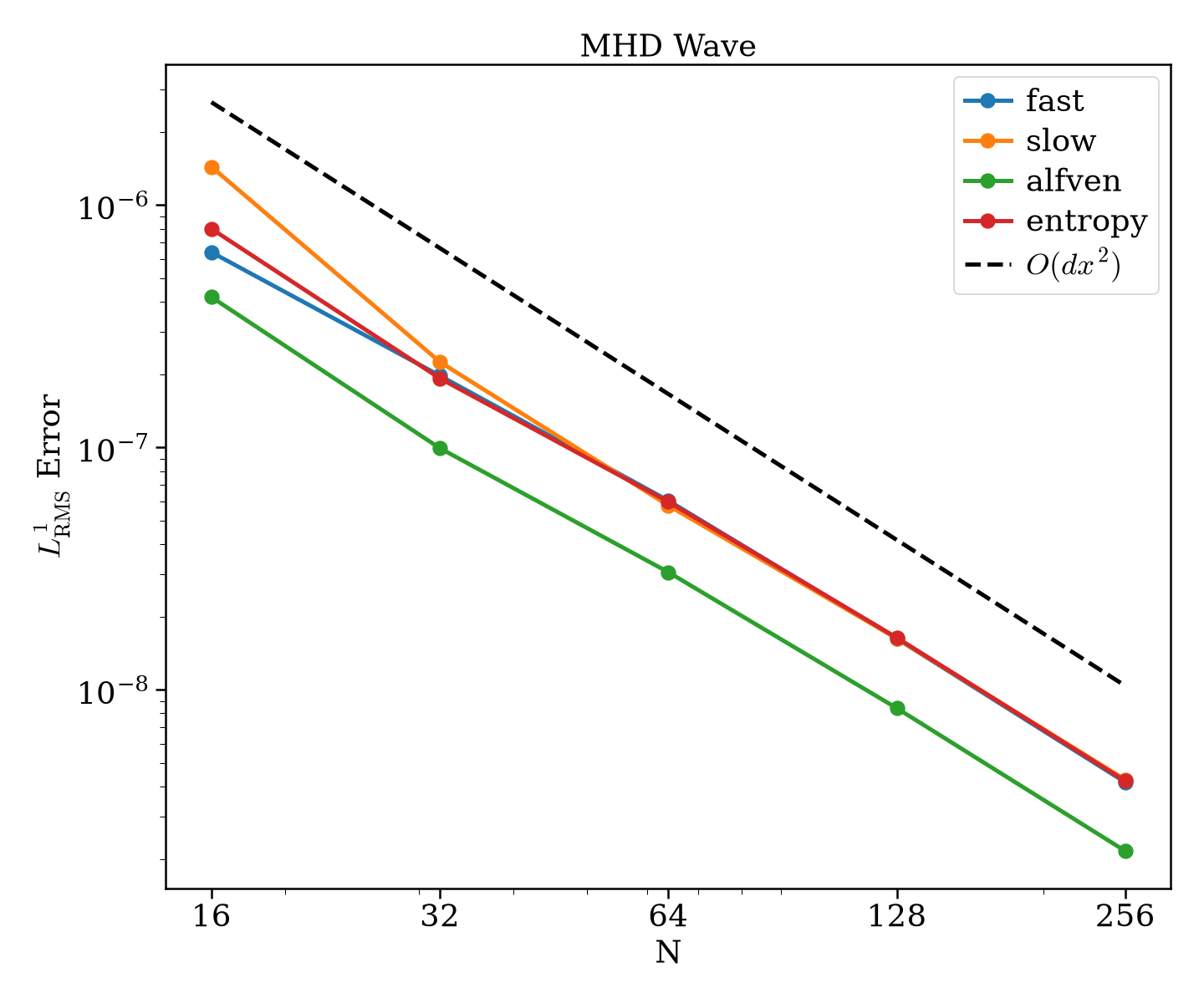}
  \includegraphics[width=0.96\columnwidth]{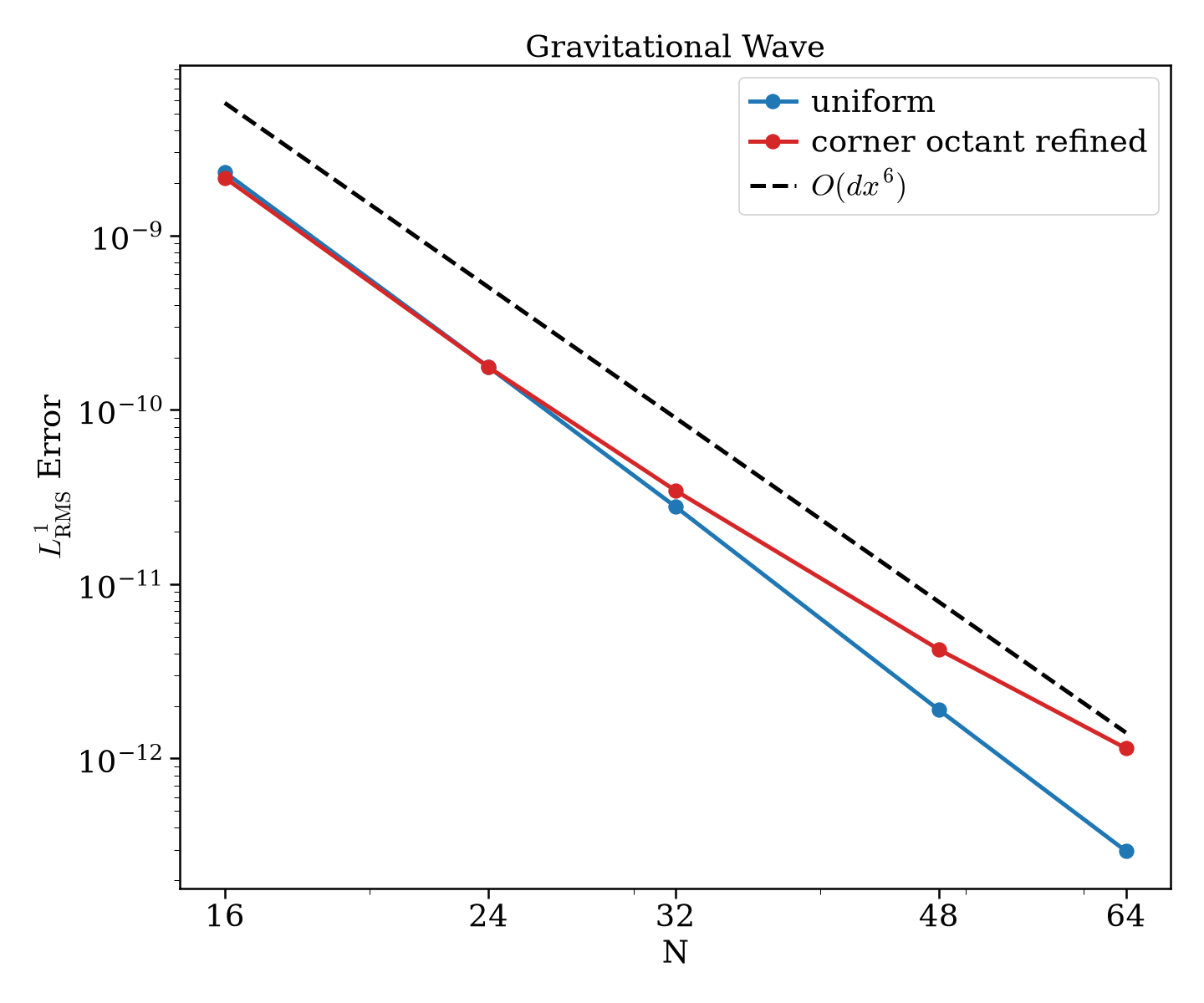}
  \caption{Code's convergence behavior on linear waves. The top panel shows the L1 errors for the fast, slow, Alfv\'en, and entropy linear MHD
  waves at different resolutions, while the bottom panel shows that for gravitational waves. In both panels, the code converges at the expected spatial order.}
  \label{fig:linear-mhd}
\end{figure}

The first test checks the spatial operators' accuracy on smooth linear waves. The relativistic
MHD wave tests include the fast, slow, Alfv\'en, and entropy modes. 
The background has $\rho_0=1$, a sound speed $c_s=10^{-2}c$, and a
gamma-law equation of state with $\Gamma=5/3$.  The background
magnetic field is $(B^1,B^2,B^3)=(1,\sqrt{2},0.5)$ and the perturbation
amplitude is $\delta\rho/\rho_0=10^{-5}$ or $\delta v_{\rm trans}=10^{-5}c_s$. The entropy mode has zero
a background advection velocity $v^1_0=10^{-2}c$, while the other modes are
initialized at rest.  The domain is $[0,1]\times[0,0.5]\times[0,0.5]$, with waves moving along the $x$ direction. 
For these tests, we use the piecewise-linear reconstruction method with the MC limiter on a uniform mesh, with
resolution ladder spans 16, 32, 64, 128, and 256 cells per wavelength. We evolve the waves for
one cycle of propagation and then measure the L1 error compared to the initial state.
The top panel of Figure~\ref{fig:linear-mhd} shows the error of these MHD waves against a second-order
reference slope.  All four modes decrease at approximately the expected
$\mathcal{O}(\Delta x^2)$ rate over the plotted range.

The corresponding Z4c linear wave test evolves a smooth gravitational wave on both a
uniform mesh and a mesh with a refined corner octant, following \citet{zhu2025}. 
The initial data are a transverse-traceless plane wave with amplitude
$6\times10^{-8}$ and three wavelengths across the unit cube, propagating
along $(1,1,1)/\sqrt{3}$. We use the 6th order spatial differencing scheme,
with the fifth order prolongation/restriction scheme. The L1 errors are measured after one
wave propagation cycle. Both uniform and refined sequences follow
the sixth-order reference trend in the bottom panel of Figure~\ref{fig:linear-mhd}. At higher resolution,
the corner-refined sequence gradually switches to a fifth order slope, which means that the prolongation/restriction error 
starts to dominate. This is consistent to that found in \citet{zhu2025} and at typical resolution of production runs, the
prolongation/restriction error is significantly lower than spatial differencing errors.

\subsection{Relativistic Kelvin-Helmholtz Instability}
The Kelvin-Helmholtz instability (KHI) is a standard test for (relativistic) hydrodynamic codes (e.g., \citet{mignone2012,Stone2020}). The relativistic KHI
in {\sc Chana} is setup as a two-dimensional SRMHD problem with a $[0,1]\times[-0.25,0.25]$ domain, 
resolved by $128\times64$ root cells and two refinement levels. The effective resolution is thus $512\times256$ cells. The medium is initially set to have $\rho=1$, $p=20$, and
$\Gamma=4/3$, with a uniform field $(B^1,B^2,B^3)=(\sqrt{2/5},0,10\sqrt{2/5})$. The shear is
$v^1=\tanh(100y)/4$ and the seed perturbation is $v^2=\sin(2\pi x)\exp(-100y^2)/400$.  
The piecewise linear method with the MC limiter is used, together with both the HLLE and HLLD Riemann solvers.

\begin{figure}
  \centering
  \includegraphics[width=0.96\columnwidth]{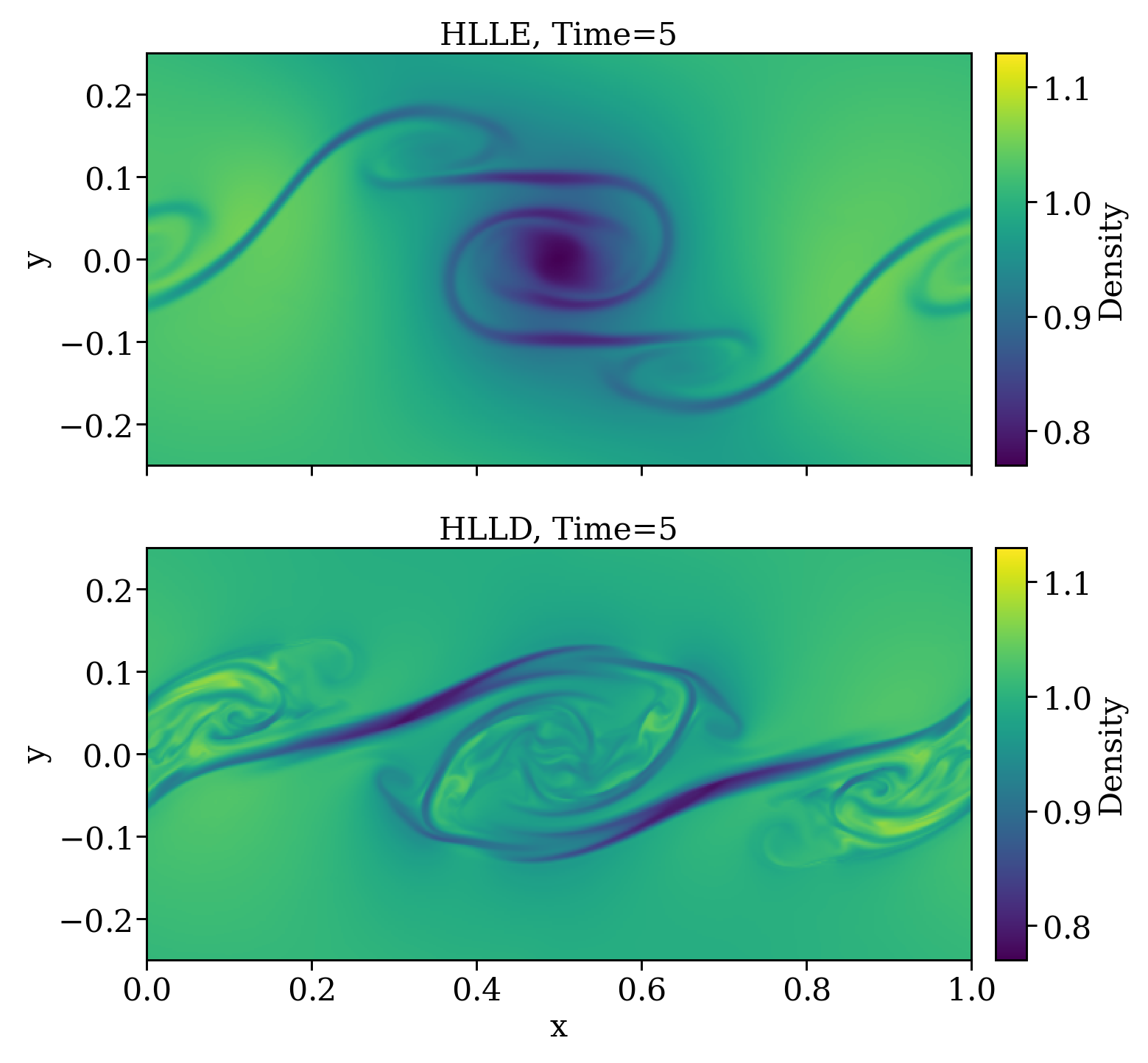}
  \caption{Rest-mass density in the relativistic Kelvin-Helmholtz test at
  $t=5$ using HLLE and HLLD Riemann solvers. The expected pair of rolled-up shear layers and a central vortex
  are developed in both cases, while the HLLD solver is significantly less diffusive than the HLLE solver.}
  \label{fig:sr-kh}
\end{figure}

Figure~\ref{fig:sr-kh} shows the results at $t=5$ for both Riemann solvers. The relativistic Kelvin-Helmholtz calculation has developed the
expected pair of rolled-up shear layers and a central vortex in both cases. The HLLE case reproduces a structure very similar to that presented
in \citet{mignone2012}, even if the meshes are quite different. This is possibly because the numerical diffusion suppresses the finer
structures introduced by the coarse-fine interfaces. The HLLD case with the same spatial resolution shows much finer structures,
which is expected because the HLLD solver is less diffusive. 

\subsection{Relativistic Magnetized Blast Wave}
The evolution of a magnetized blast wave moving at relativistic speeds provides 
a further test of the relativistic MHD module in the code. 
A two-dimensional case is considered: 
inside $r=\sqrt{x^2+y^2}=0.8$ the density and pressure are $(10^{-2},1)$,
outside $r=1$ they are $(10^{-4},5\times10^{-3})$, and both quantities vary
linearly across $0.8<r<1$.  The equation of state has $\Gamma=4/3$ and the
uniform field is $(B^1,B^2,B^3)=(1/\sqrt{20},1/\sqrt{20},0)$.  The uniform
run covers $[-6,6]^2$ with $512^2$ in-plane cells. 
The AMR run uses the same domain and the block size is chosen to be 8 to build a more flexible 
AMR hierarchy. It starts with a $4\times4$ root mesh and adds four refinement levels above
with a L\"ohner density indicator (refine/coarsen thresholds 0.02/0.005),
reaching the same finest spacing. Both calculations evolved to $t=4$ with a CFL 
number of 0.3.

\begin{figure}
  \centering
  \includegraphics[width=0.96\columnwidth]{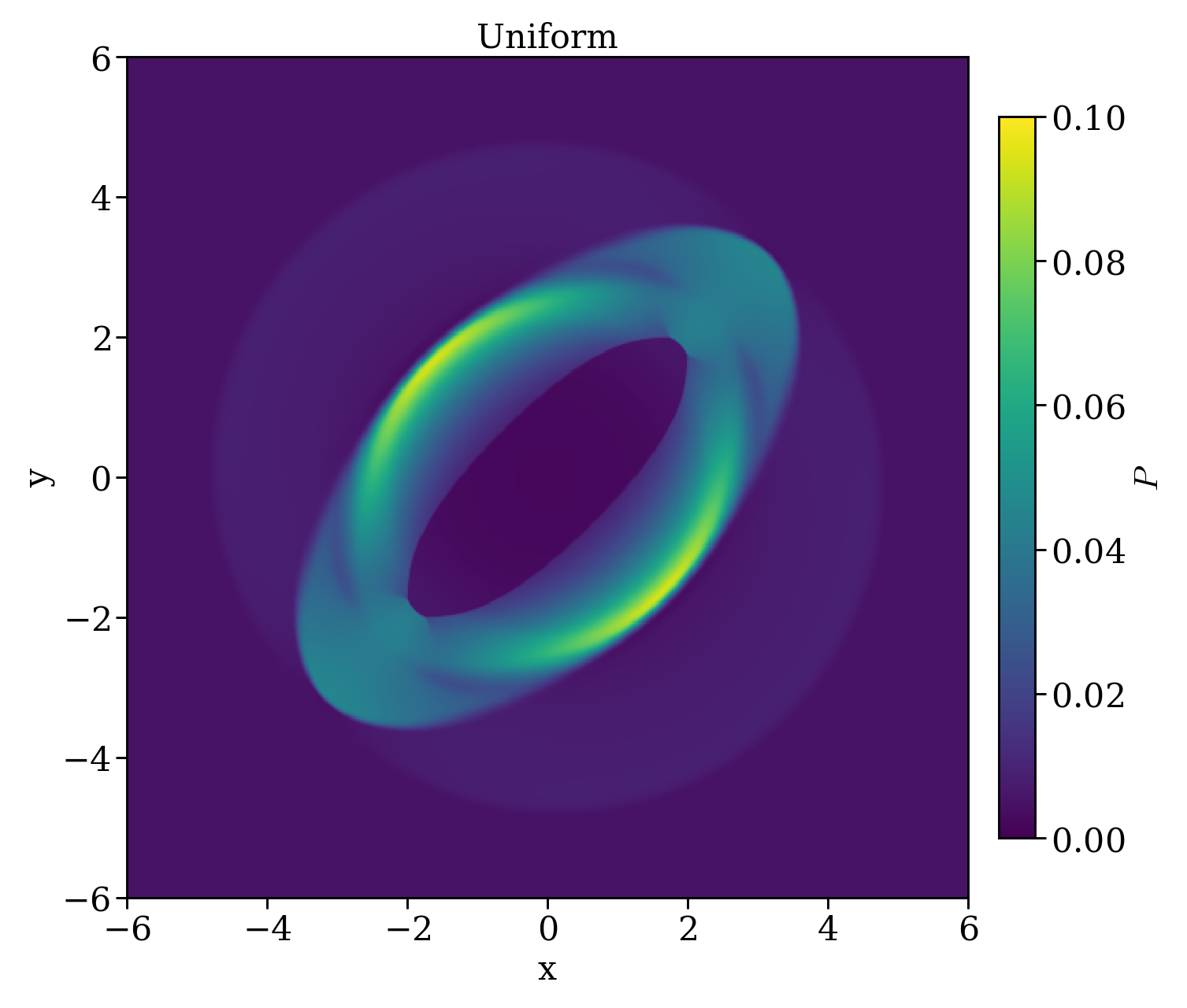}
  \includegraphics[width=0.96\columnwidth]{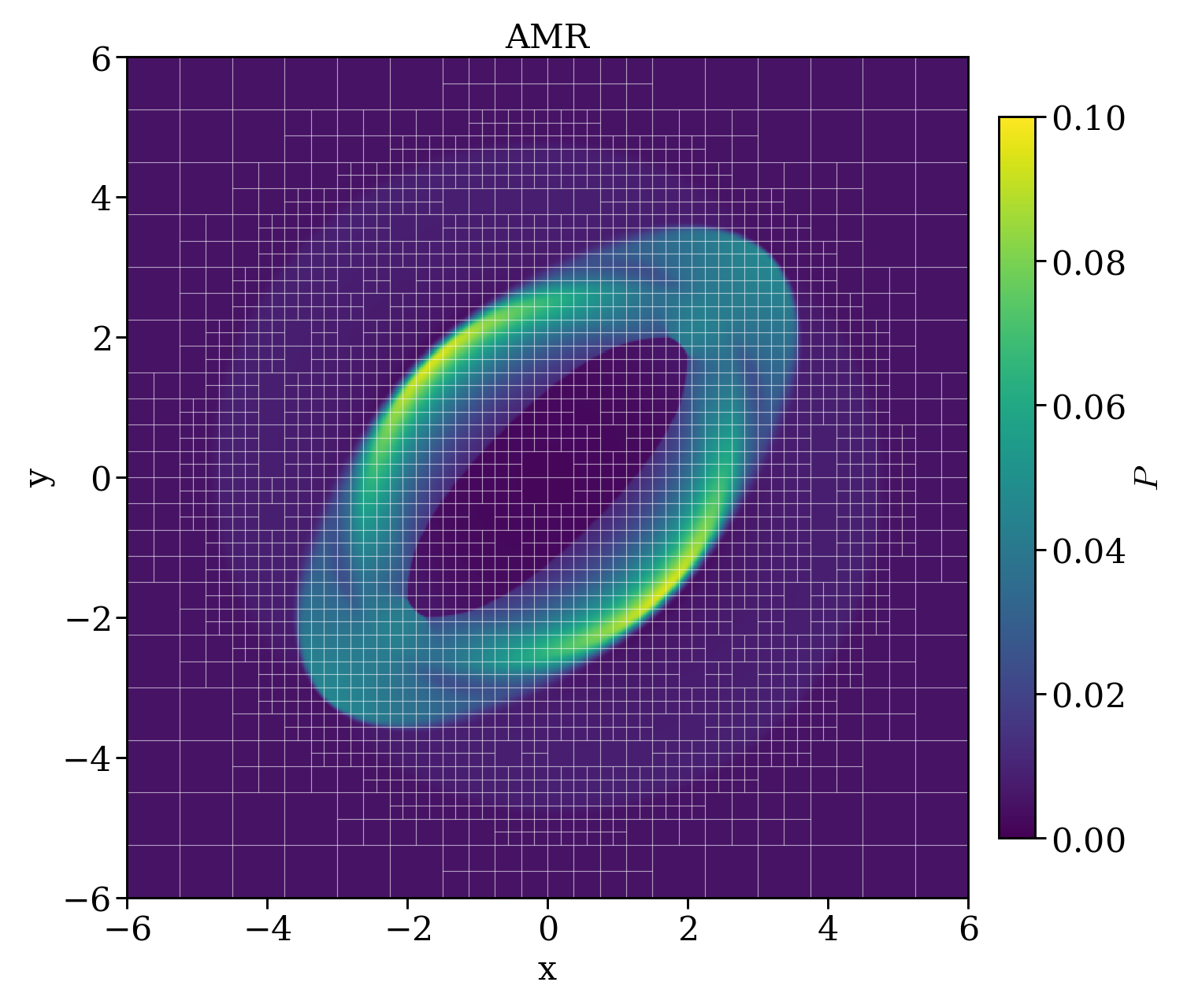}
  \includegraphics[width=0.96\columnwidth]{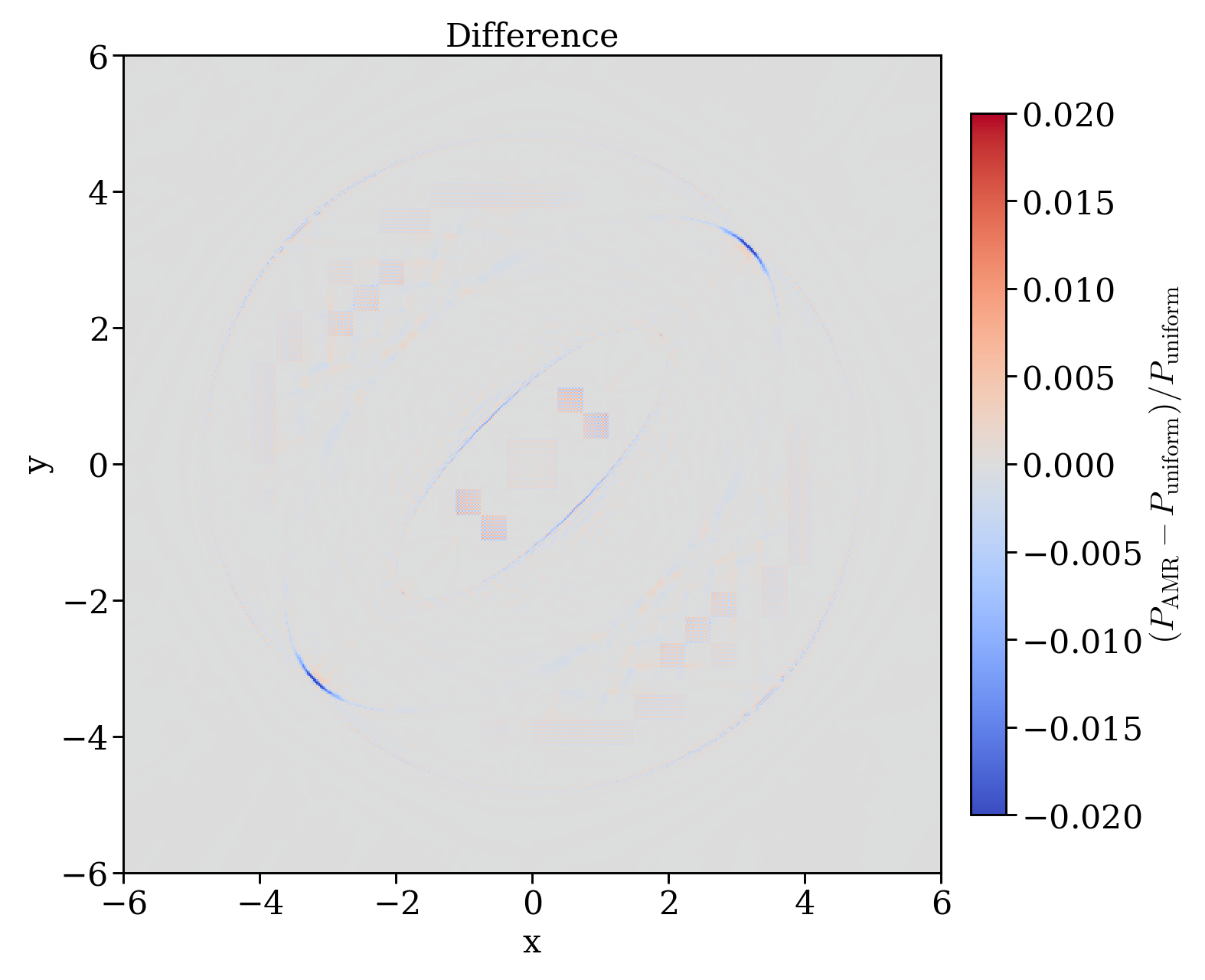}
  \caption{Pressure in the relativistic magnetized blast wave. The top panel shows the results on a uniform
  mesh and the middle panel shows the corresponding AMR results at the same finest spacing, together with the AMR hierarchy. 
  The bottom panel shows the fractional differences between the uniform and AMR tests. The maximum fractional difference between uniform and AMR cases 
  is always below two percents, and the RMS fractional error is 7.1$\times10^{-4}$. }
  \label{fig:blast-pressure}
\end{figure}

Figure~\ref{fig:blast-pressure} shows the results at the end of the simulations.
The magnetized blast produces an expanding and strongly anisotropic pressure
shell, which is recovered by both the uniform and AMR runs. 
The maximum fractional difference between uniform and AMR cases is below 2\% throughout the domain at $t=4$, and 
the RMS fractional difference is 7.1$\times10^{-4}$.

\subsection{Diffusion Limit in a Moving Medium}
\label{sec:moving_diffusion}
The first test on the radiation transport module focuses on the diffusion limit. Since {\sc Chana} by default
uses the Lorentz covariant Levermore closure in the lab-frame instead of the fluid-frame, 
it's important to test if it can reproduce the correct diffusion limit in a relativistically moving medium.

This test uses one grey group, a uniform medium moving at $0.5$ the speed of light, and pure scattering with $\sigma=1000$.  
The initial radiation energy is a Gaussian $E(x)=\exp(-9x^2)$, 
with its flux set to the fully trapped comoving value so that the pulse translates with the fluid from the beginning \citep{radice2022}.  
The spatial resolution is set to be $\Delta x=0.01$. 

\begin{figure}
  \centering
  \includegraphics[width=0.96\columnwidth]{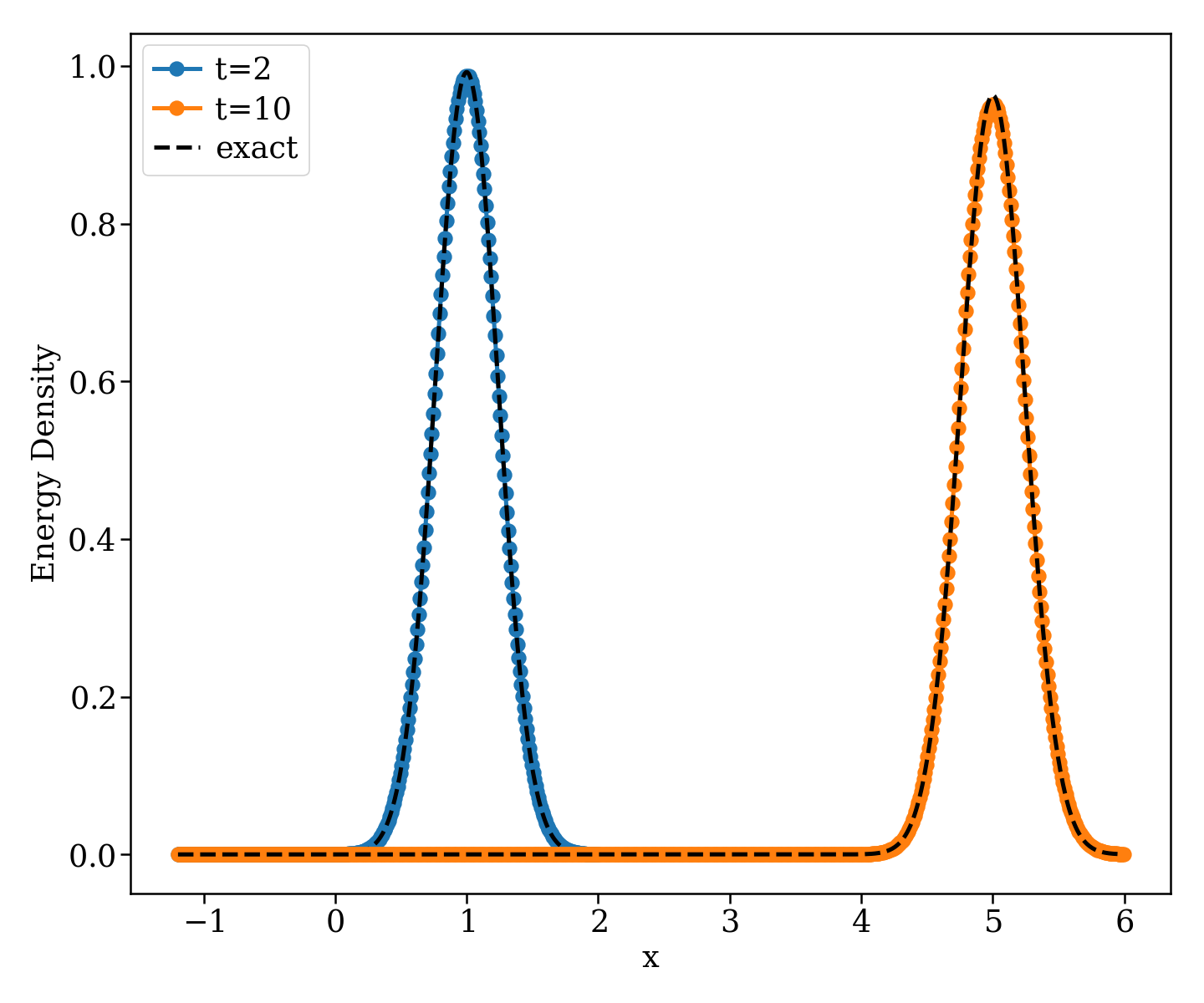}
  \caption{Moving diffusion test at $t=2$ and $t=10$.  The analytic solution is shown as black dashed lines, while the
  markers depict the numerical result. At both times, the numerical energy-density profile lies on the analytic translated and diffused solution over the entire domain. }
  \label{fig:moving-diffusion}
\end{figure}

Figure~\ref{fig:moving-diffusion} shows the test results at $t=2$ and $t=10$. 
At both times, the numerical energy-density profile lies on the analytic translated and diffused solution over the entire domain. 
This means that the flux treatment described in Section~\ref{sec:num-m1} together with
the Levermore closure evaluated in the Eulerian frame is able to give the correct diffusion-limit behavior in this case. \citet{radice2022} has also discussed that the full radiation-matter coupling treatment is required to capture the behavior of the trapped radiation, while approximate solvers may give wrong advection speeds or rates of diffusion. Therefore, this test also partly validates {\sc Chana}'s radiation-matter coupling implementation. Although candidate solutions are proposed by approximate solvers, they are accepted only if the convergence criterion based on the exact residual function is satisfied.

\subsection{Doppler and Gravitational Shifts}
\label{sec:test-shifts}
The two cases shown in this subsection test the frequency-space operator 
with an emitting sphere whose radiation propagates through a transparent medium, following \citet{Anninos2020}.  
For a more straightforward comparison to the same test in that paper, units and physical constants are put back in the 
following description.

The sphere has $R_*=10\,\mathrm{km}$ and emits a $5\,\mathrm{MeV}$ Fermi-Dirac spectrum
sampled by 30 logarithmic groups between 1 and 50 MeV.  The radial domain is
$[0,400\,\mathrm{km}]$, represented by 400 cells, and the final time
is two light-crossing times.

For the Doppler test the metric is flat and the prescribed infall velocity is
zero for $r\leq 70\,\mathrm{km}$, decreases linearly to $-0.2c$ at
$r=80\,\mathrm{km}$, and follows
$v(r)=-0.2c(80\,\mathrm{km}/r)^2$ further out. For the gravitational
test the matter is at rest and the metric is generated by a uniform sphere of
density $\rho_*=9\times10^{14}\,\mathrm{g\,cm^{-3}}$.  With
$M=4\pi\rho_*R_*^3/3$, the prescribed potential and lapse are
\begin{equation}
  \begin{aligned}
    \Phi(r)&=
    \begin{cases}
      -GM(3R_*^2-r^2)/(2R_*^3), & r\leq R_*,\\
      -GM/r, & r>R_*,
    \end{cases}\\
    \alpha(r)&=\sqrt{1+2\Phi(r)/c^2}.
  \end{aligned}
  \label{eq:test-shift-potential}
\end{equation}
The exact mean energy outside the emitting region follows from conservation of
the radiation energy along a stationary characteristic,
\begin{equation}
  \begin{aligned}
    \frac{\langle\varepsilon\rangle(r)}{\langle\varepsilon\rangle_0}&=
      \frac{\alpha_{0}}
           {\alpha(r)W(r)[1+v(r)/c]},\\
    W(r)&=\left[1-v(r)^2/c^2\right]^{-1/2},
  \end{aligned}
  \label{eq:test-shift-law}
\end{equation}
where $\langle\varepsilon\rangle_0$ is the discrete mean of the emitted
groups.  The Doppler test has $\alpha=1$, while the gravitational test has
$v=0$.  

\begin{figure}
  \centering
  \includegraphics[width=0.96\columnwidth]{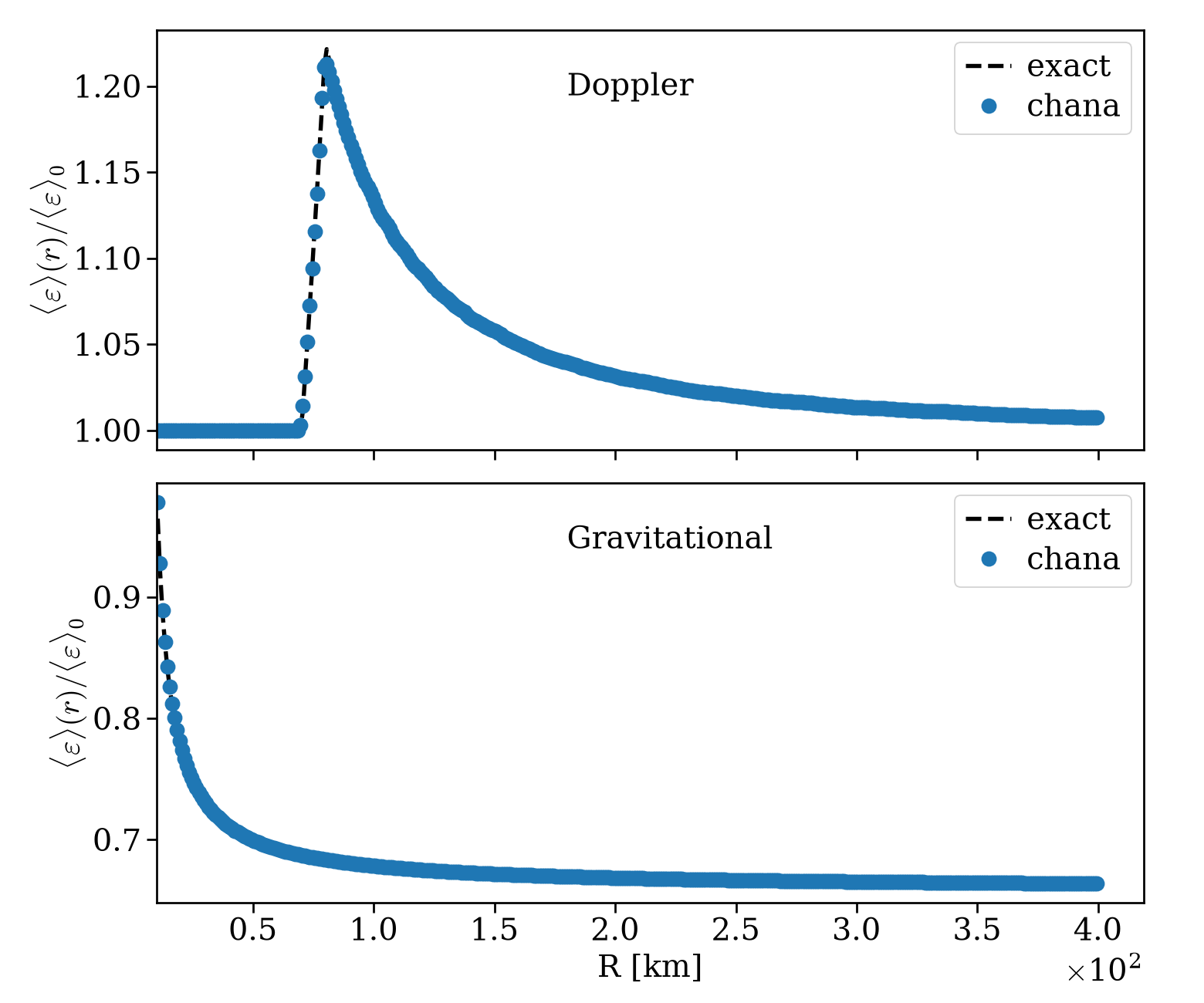}
  \caption{Mean radiation energy outside the emitting sphere for the Doppler
  (top) and gravitational-redshift (bottom) tests.  The blue points are the
  numerical group-weighted means and the black dashed curves are the exact
  stationary-shift laws given by Eq.~\eqref{eq:test-shift-law}. The good agreement
  between numerical results and the exact solutions shows that the frequency-space drift
  treatment described in Section~\ref{sec:num-m1} works accurately. }
  \label{fig:shift-tests}
\end{figure}

The final profiles in Figure~\ref{fig:shift-tests} show the simulated
group-weighted mean energy against Eq.~\eqref{eq:test-shift-law}.  The Doppler
profile follows the prescribed blueshift through the velocity ramp and returns
toward the emitted value as the infall speed decreases outside the ramp, while the
gravitational profile decreases monotonically with radius. The good agreement
between the numerical results and the exact solutions shows that the frequency-space drift
treatment described in Section~\ref{sec:num-m1} works accurately.

\subsection{Cylinder Shadow}
The first multi-dimension test on the radiation module is a cylinder shadow test,
in which an initially uniform radiation field is attenuated by an absorbing cylinder.
This is an one-group M1 calculation on a frozen matter reservoir in flat
spacetime.
The domain is $[-2,4]\times[-2,2]$, resolved by
$480\times320$ cells. The cylinder has radius one, absorption opacity 
$\kappa=1$ and zero scattering inside, giving a
central optical depth of two. Medium outside the cylinder is set to $\kappa=10^{-10}$.
Radiation starts at a negligible floor and a
unit, forward-peaked beam is injected through the left $x$ boundary.

\begin{figure}
  \centering
  \includegraphics[width=0.96\columnwidth]{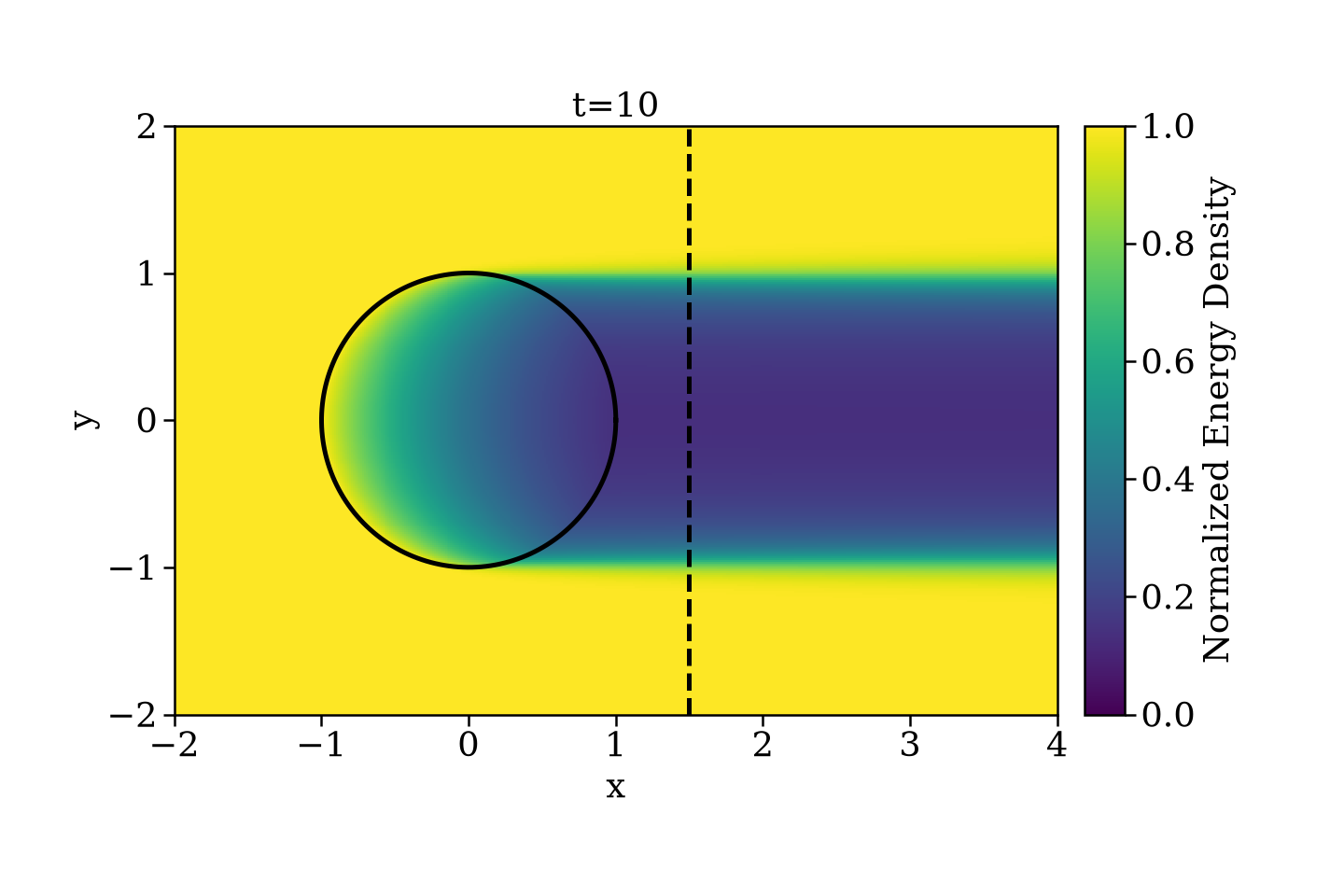}
  \includegraphics[width=0.80\columnwidth]{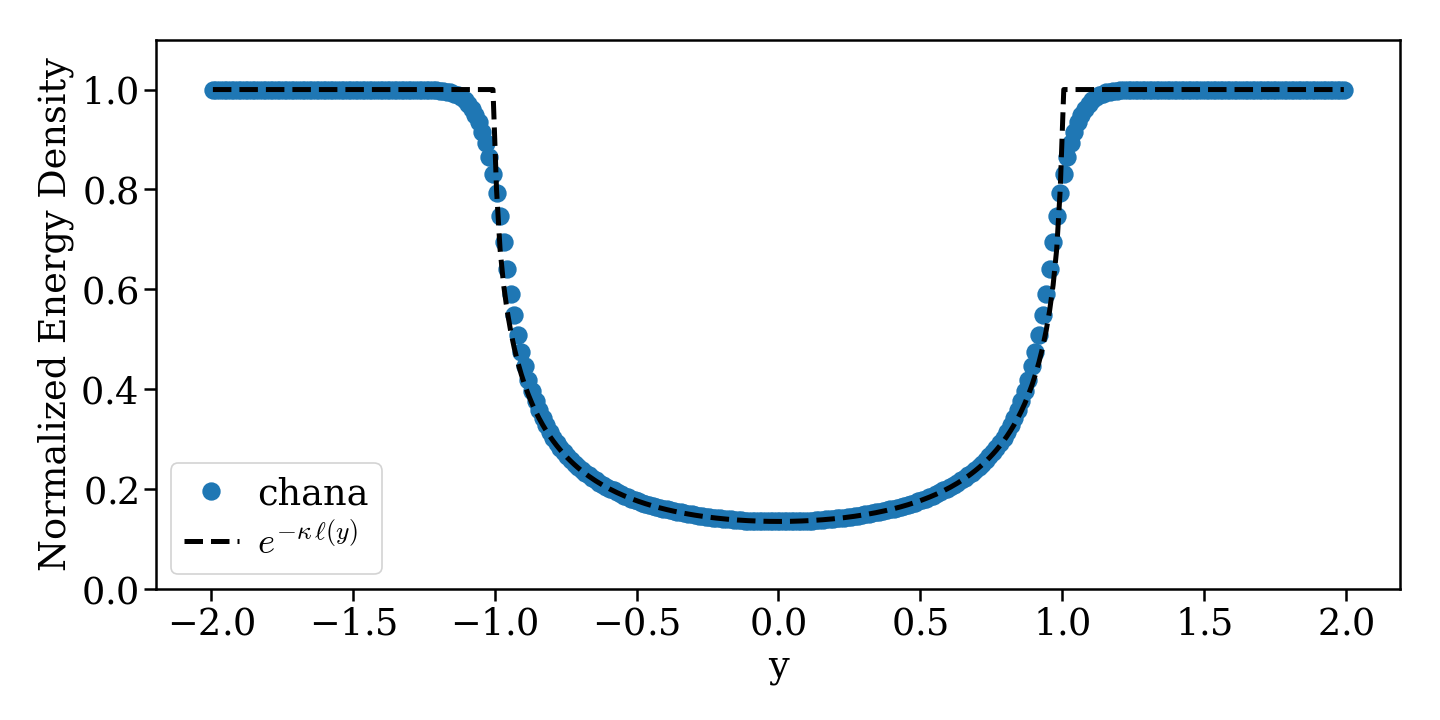}
  \caption{Results of the cylindrical shadow test at $t=10$. The top panel shows the spatial distribution of
  radiation energy density with the cylinder boundary marked by a circle. The bottom panel compares the numerical
  energy distribution on a transverse probe (indicated by the black dashed line in the top panel) with the 
  analytical attenuation. The numerical profile follows the analytic solution closely, while some discrepancies can be seen
  at the smoothed shadow edges. The mismatch is caused by the slightly over-diffusive spatial flux solver.}
  \label{fig:shadow}
\end{figure}

The $t=10$ energy-density map in Figure~\ref{fig:shadow} shows the persistent
low-density shadow downstream of the obstacle.  On a transverse probe, the
numerical profile follows the analytic attenuation
$\exp[-\kappa\ell(y)]$ closely, while the visible discrepancy is confined
to the smoothed shadow edges. This shows that the second-order spatial flux is slightly over-diffusive, 
but given its stability and simplicity we still use it as the default choice. 

\subsection{Radiation Beam Near a Black Hole}
\label{sec:test-bhbeam}
The next test focuses on the spacetime curvature terms in {\sc Chana}.
This is a five-group, zero-opacity M1 calculation on a frozen Schwarzschild
metric in Kerr-Schild coordinates. The domain is
$[0,5M]\times[0,4M]$, with $M=1$ in code units, and is resolved by
$200\times160$ cells ($\Delta x=\Delta z=0.025M$). The initial radiation is
a low-amplitude background, while a beam of half-width $0.25M$ and unit
energy is injected at $(x,z)=(0,3.5M)$ along $+x$ through the left boundary.
Since the opacity is identically zero and the metric is held fixed, 
each group follows the same null-geodesic problem, and the reference curves are the corresponding analytic Kerr-Schild
characteristics. The test uses a CFL number of 0.3 and is evolved to $t=20M$.

\begin{figure}
  \centering
  \includegraphics[width=0.96\columnwidth]{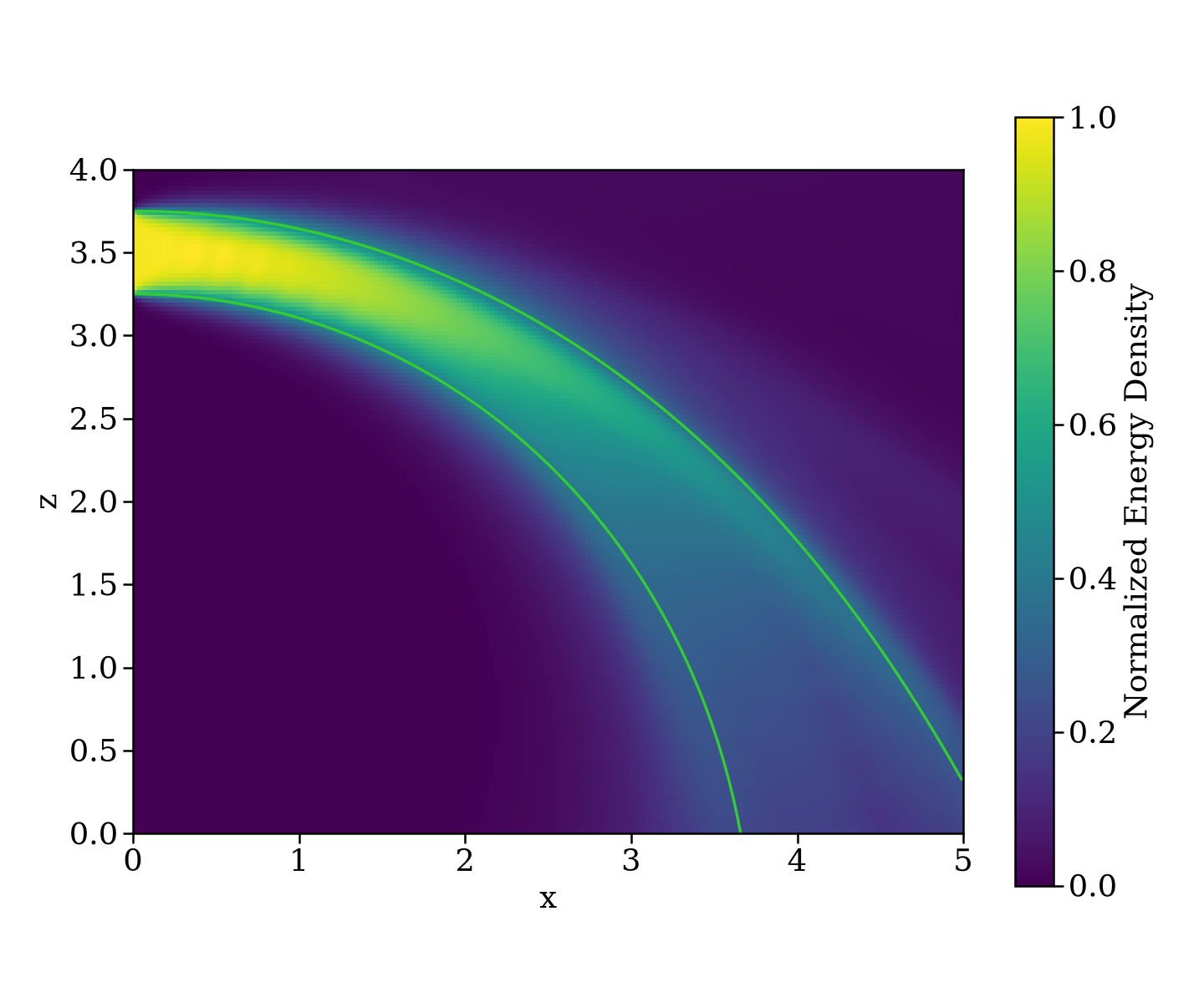}
  \caption{Normalized radiation energy density for the beam propagating near
  a black hole. The green curves mark the reference characteristic paths, and the radiation energy remains concentrated between them.
  The front broadens gradually due to the over-diffusive spatial radiation flux, but the trajectory remains aligned with the
curved reference paths across the plotted domain.}
  \label{fig:bh-beam}
\end{figure}

The normalized energy density in Figure~\ref{fig:bh-beam} remains concentrated between the two reference
characteristics as the beam bends toward increasing $x$ and decreasing $z$.
The front broadens gradually due to the over-diffusive spatial radiation flux, but the trajectory remains aligned with the
curved reference paths across the plotted domain. This broadening is due to the more diffusive spatial flux, and the results closely
reproduce that in \citet{radice2022} which uses the same method.

\subsection{Evolution of Binary Punctures}
A binary puncture test is presented to demonstrate {\sc Chana}'s capability 
to simulate dynamic spacetimes. We use the total ADM mass $M$ as the unit of
length and time. The initial data contain two non-spinning punctures with bare
masses $0.488479M$ at $(\pm6.10679,0,0)M$, with equal and opposite momenta
$(-5.10846\times10^{-4},8.41746\times10^{-2},0)M$ and
$(5.10846\times10^{-4},-8.41746\times10^{-2},0)M$. The
computational cube is $[-192,192]^3M$, with seven root blocks per
direction and eight refinement levels. Each block has eight cells along each dimension, which
is the smallest allowed block size with the 6th order spatial differencing method. This
small block size leads to a more flexible AMR hierarchy and thus add extra pressure to the 
AMR and halo exchange modules. The $\dd\chi$ criterion is used with refine/coarsen thresholds set to 0.03 and 0.005,
respectively. A shift damping of $\eta=2.10M^{-1}$ and constraint damping
$\kappa_1=0.06M^{-1}$ are used, with the K.O. dissipation parameter set to be 0.5.
The test is evolved to $t=120M$.

\begin{figure}
  \centering
  \includegraphics[width=0.96\columnwidth]{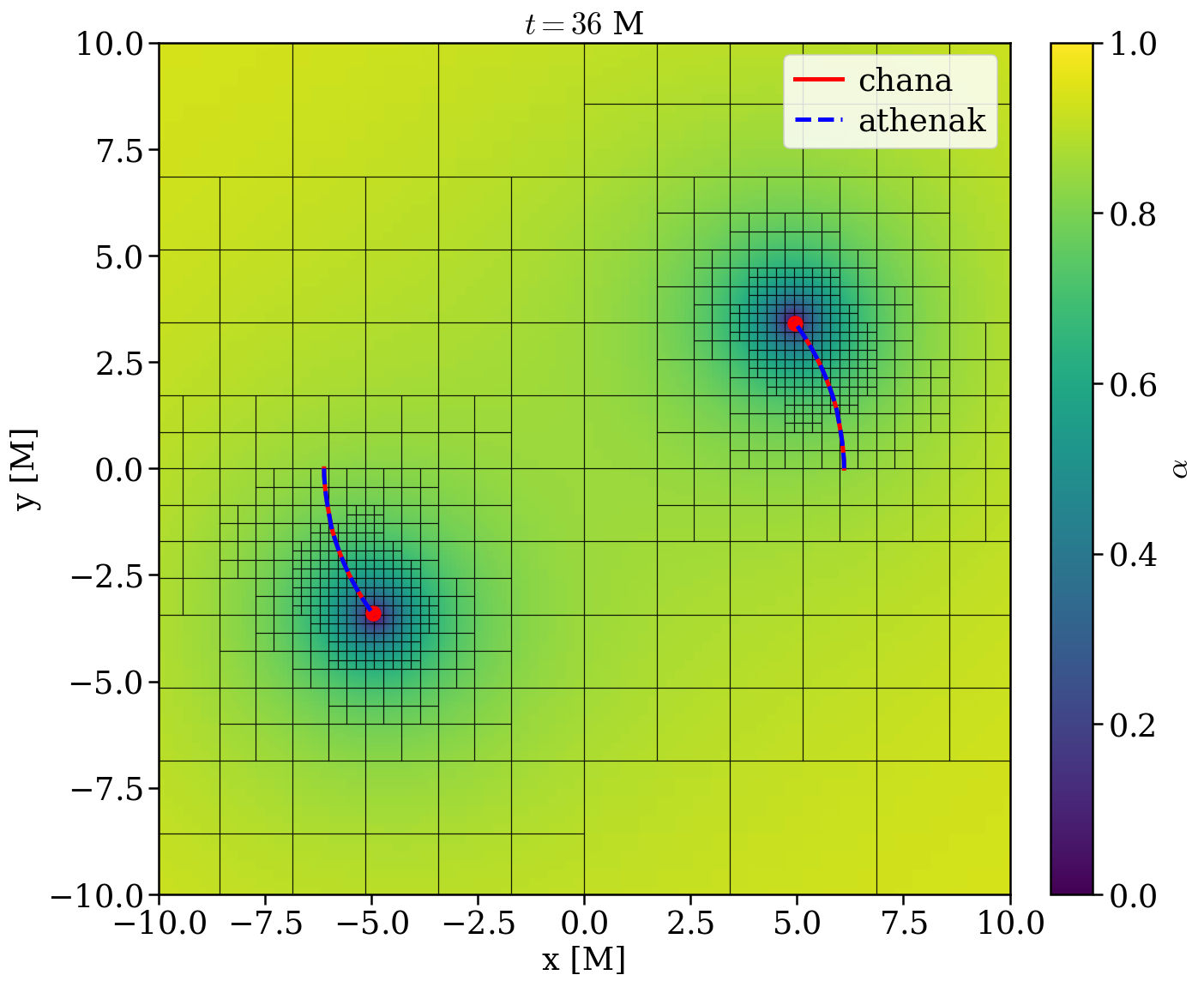}
  \caption{Results of the binary-puncture test at $t=36M$.  The color shows the lapse,
  the black mesh lines show the AMR hierarchy, and the cyan curves show the
  tracked puncture trajectories. The asymmetric block distribution around the punctures is caused by the imbalance between the refine and coarsen criteria.
  The puncture trajectories are in good agreement with the results calculated by AthenaK \citep{zhu2025}.}
  \label{fig:binary-puncture-evolution}
\end{figure}

Figure~\ref{fig:binary-puncture-evolution} shows the lapse and refinement
pattern at $t=36M$. The tracked punctures remain resolved by the nested refinement levels. The 
imbalance between the refine and coarsen criteria leads to the asymmetric block distribution
around the punctures, because the coarsening is delayed compared to refining as the punctures move. 
The puncture trajectories are in good agreement with the results calculated by AthenaK \citep{zhu2025},
despite of the very different AMR hierarchy and block sizes\footnote{AthenaK is significantly slowed down with size-8 blocks, thus
size-16 blocks are used. The min-$\chi$ AMR criterion is used in the AthenaK comparison so that it matches the setups in \citet{zhu2025}.}.

\subsection{Collapse of A Neutron Star}
Finally, the code's dynamic spacetime capability in a non-vacuum spacetime is validated by a neutron star collapse test.
The initial star is a $\Gamma=2$, $K=100$ polytrope (in units of $G=c=M_\odot=1$),
with a central density of $\rho_c=1.895\times10^{15}\,\mathrm{g\,cm^{-3}}$, just below that of the maximum-mass configuration.
Solving the TOV equations gives a gravitational mass of $M=1.637\,M_\odot$, an areal radius of $11.39\,\mathrm{km}$, and a baryonic mass of $1.798\,M_\odot$.
Outside the star is an atmosphere of $\rho=10^6\,\mathrm{g\,cm^{-3}}$ on the same polytrope, and the matter is evolved with a $\Gamma=2$ ideal-gas equation of state.
The initial spacetime is the TOV metric in isotropic coordinates, i.e., conformally flat with zero shift and zero extrinsic curvature,
and the matter is cell-averaged onto the mesh with a $3\times3\times3$ Gauss-Legendre quadrature. The collapse is triggered by reducing the pressure and internal energy
of the matter by 5\% after the hydrostatic initialization solve, which leaves the star with more baryonic mass than any equilibrium of the depleted polytrope can support.
The metric is resolved after this reduction using the method described in \citet{Cordero-Carrion2009} and total mass-energy of the initial data is then $1.628\,M_\odot$.

The computational domain is $[-128,128]^3\,\mathrm{km}$, covered by four root blocks per direction, each with $32^3$ cells.
Three levels of refinement are placed around the star, so the finest spacing is $0.25\,\mathrm{km}$ over $|x^i|\leq16\,\mathrm{km}$. The $\dd\chi$ indicator with a refine threshold of 0.02 is used. With $M$ as the unit, the shift damping is $\eta=1.69M^{-1}$ and the constraint damping is $\kappa_1=0.048M^{-1}$ with $\kappa_2=0$,
and the K.O. dissipation parameter is set to be 0.5. The piecewise linear method with the MC limiter and the HLLE Riemann solver are used for the matter, and the Z4c uses the 6th order spatial differencing scheme. The test is evolved with a CFL number of 0.3 to $t=1.0\,\mathrm{ms}\approx124M$.

\begin{figure}
  \centering
  \includegraphics[width=0.96\columnwidth]{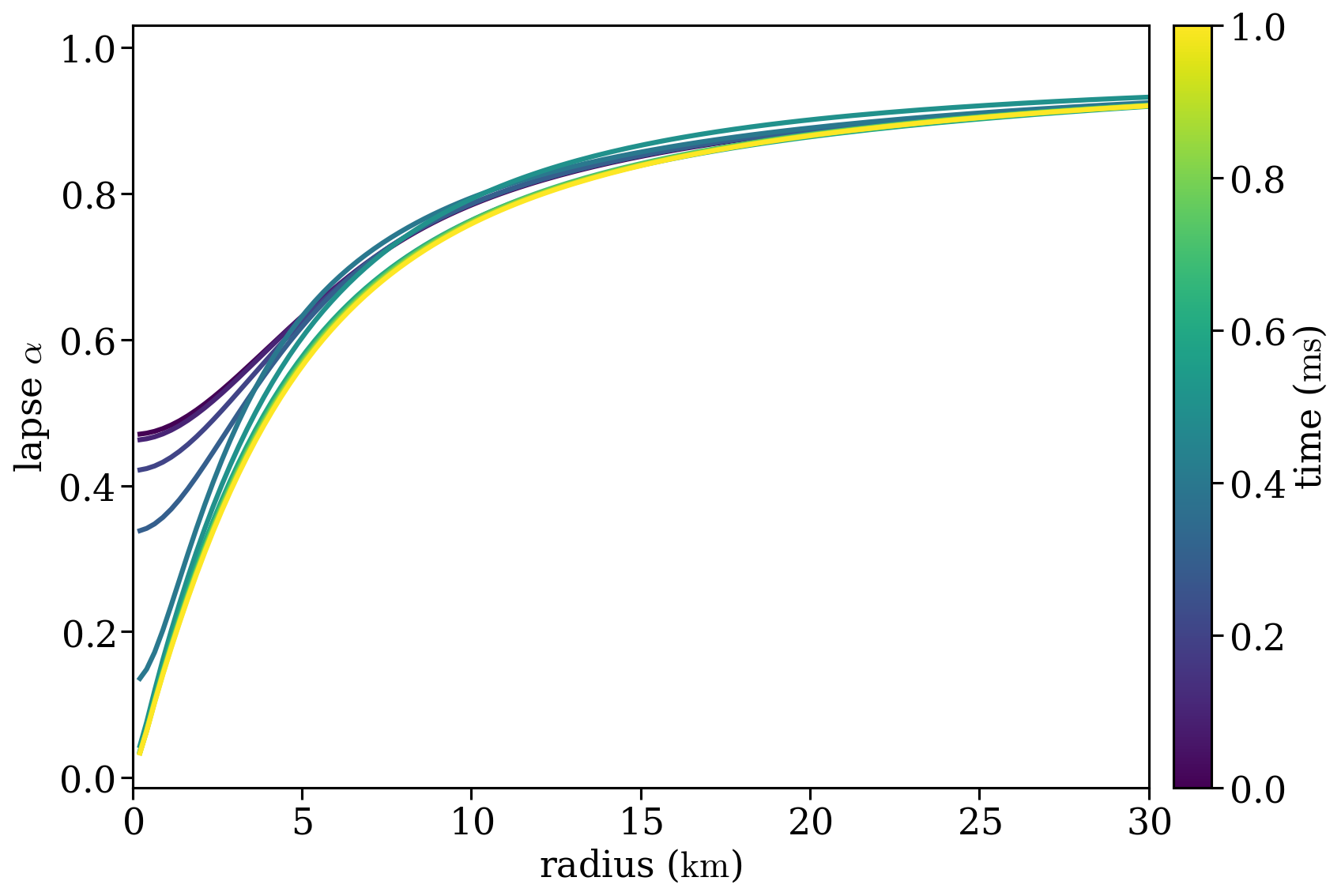}
  \includegraphics[width=0.96\columnwidth]{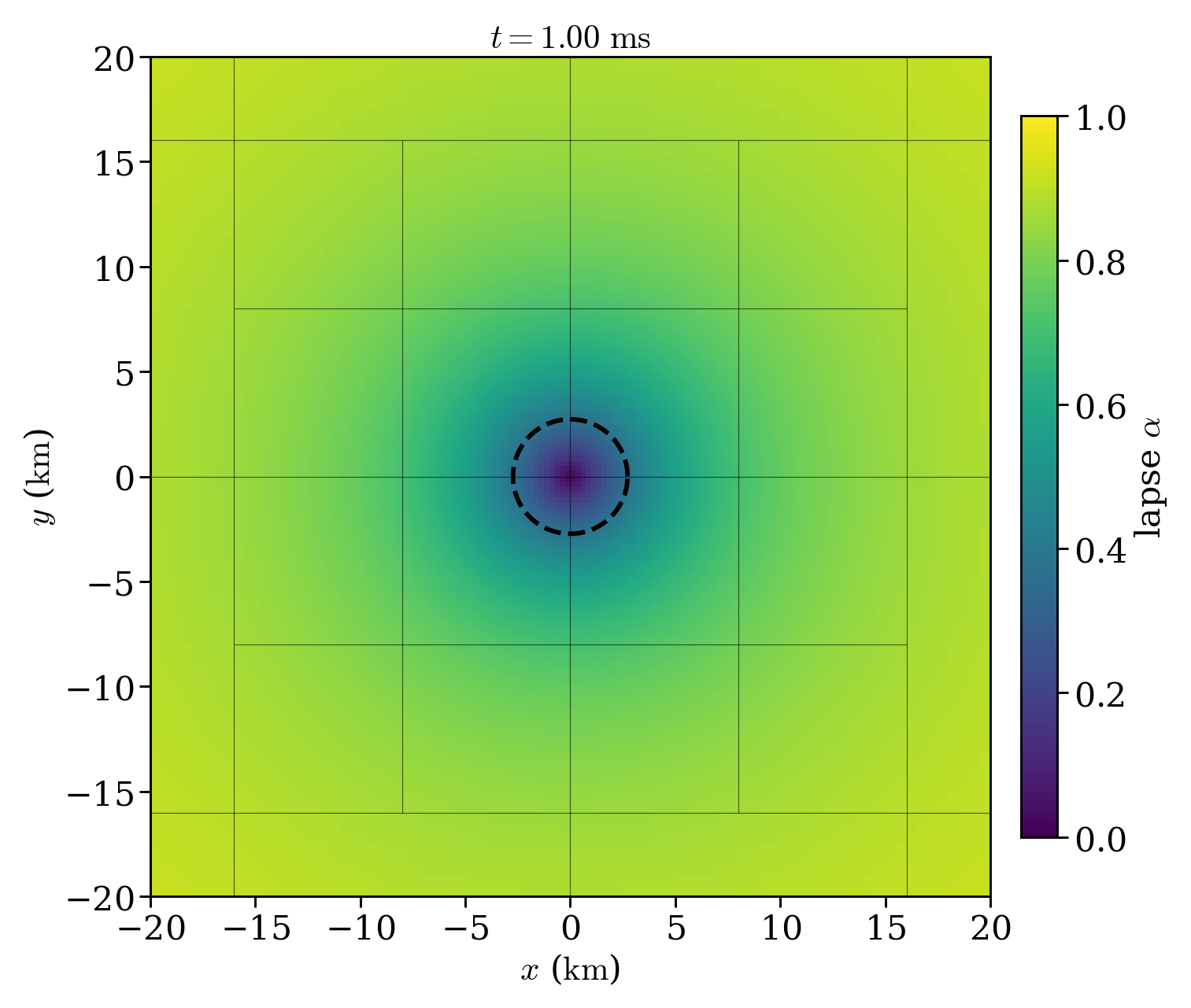}
  \caption{Results of the neutron star collapse test. The top panel shows the lapse along the $x$ axis every $0.1\,\mathrm{ms}$, colored by time.
  The profiles after $0.6\,\mathrm{ms}$ nearly overlap each other, showing that the spacetime has settled into a stationary slice of a black hole.
  The bottom panel shows the lapse in the $z=0$ plane at $t=1\,\mathrm{ms}$, with the black mesh lines showing the AMR hierarchy and the black dashed circle
  marking the apparent horizon, whose areal radius gives a black hole mass of $1.630\,M_\odot$.}
  \label{fig:ns_collapse}
\end{figure}

Figure~\ref{fig:ns_collapse} shows results of this test. The top panel shows the lapse profile along the $x$-axis at different times, while the bottom panel shows a snapshot at $t=1\,\mathrm{ms}$ together with the apparent horizon and AMR hierarchy. The central lapse starts at 0.471, decreases slowly during the first $0.2\,\mathrm{ms}$, drops rapidly between 0.3 and $0.5\,\mathrm{ms}$ as the star collapses, and then settles to $\approx0.034$ at the grid point nearest to the center. The lapse profiles after $0.6\,\mathrm{ms}$ nearly overlap each other, showing that the $1+\log$ slicing and the Gamma-driver shift have brought the spacetime to the expected stationary trumpet-like slice of a black hole. 
The apparent horizon, found as the outermost sphere with zero outgoing null expansion assuming spherical symmetry, first appears at $t=0.5\,\mathrm{ms}$.
Its coordinate radius drifts slowly from 2.55 to $2.73\,\mathrm{km}$ as the gauge evolves, while its areal radius stays at $4.81\,\mathrm{km}$,
giving a black hole mass of $M_{\rm BH}=R_{\rm AH}/2=1.630\,M_\odot$. By $t=1\,\mathrm{ms}$ all of the matter is inside the horizon and the baryonic mass is conserved to better than $10^{-4}$. Since a spherical collapse emits no gravitational waves, the black hole should carry the mass-energy of the initial data, in agreement with the measured value to $0.2\%$.

\subsection{Performance}
The performance of {\sc Chana} is measured using Nvidia A100 GPUs on the Perlmutter supercomputer of National Energy Research Scientific Computing (NERSC). The setup is a core-collapse supernova case, and the initial condition is set to be an 18 $M_\odot$ model from \citet{sukhbold2018}, evolved to 10 milliseconds post-bounce so that neutrino radiation is strong. The tabulated SFHo equation of state \citep{Steiner2013} is used, while neutrino-matter interaction rates are taken from \citet{2006NuPhA.777..356B}, the same as those used in {\sc Fornax} CCSN simulations \citep{skinner2019,vartanyan2018b,burrows_2020,burrows_correlations_2024,wang_low_2024}. There are three neutrino species, each with 12 energy groups. The piecewise-linear method with the MC limiter is used for MHD and radiation variables, while the Z4c uses the 6th order spatial differencing scheme. 

The code's performance with different block sizes is measured. The number of total active cell is chosen to be 524,288=$2^{19}$, which is the same order of magnitude of the actual workload of a production simulation. Block sizes of 8, 16, 32, and 64 are tested. The results are 216k, 293k, 334k, and 171k cell-updates per second when the eSSPRK$(10,4)$ time integrator is used, and with the standard RK4 integrator the numbers become 525k, 714k, 820k, and 421k. The factor of about 2.5 is a direct consequence of the 2.5 times fewer stages per step in the standard RK4 method. The decreasing performance of blocks with 64 cells is because there are only two blocks in the tested domain and block-oriented kernels have very little coarse-grained parallelism. From this test, it can be concluded that a block size of 32 gives the best performance, while a block size of 16 leads to more AMR flexibility but slightly lower computation efficiency. In addition, increasing the number of active cells per GPU improves the GPU performance. Because the size-32 blocks have a lower face-to-bulk ratio and memory cost, each A100 GPU can handle $2^{20}=1,048,576$ cells (size-8 blocks will lead to out-of-memory), which leads to 445k eSSPRK$(10,4)$ cell-updates and 1.1M RK4 cell-updates per second. It's worth mentioning that the code's performance strongly depends on the acceptance rate of the candidate solutions estimated by the approximate radiation-matter solvers, and the acceptance rate can change significantly in the domain. In the above performance tests, the domain is refined to be the inner 100 km of the CCSN model where neutrinos and matter strongly couples, therefore the acceptance rate is lower. For a more general simulation setup, with many blocks locate at large radii and candidate solutions work well, a higher performance is expected. 

\begin{figure}
  \centering
  \includegraphics[width=0.96\columnwidth]{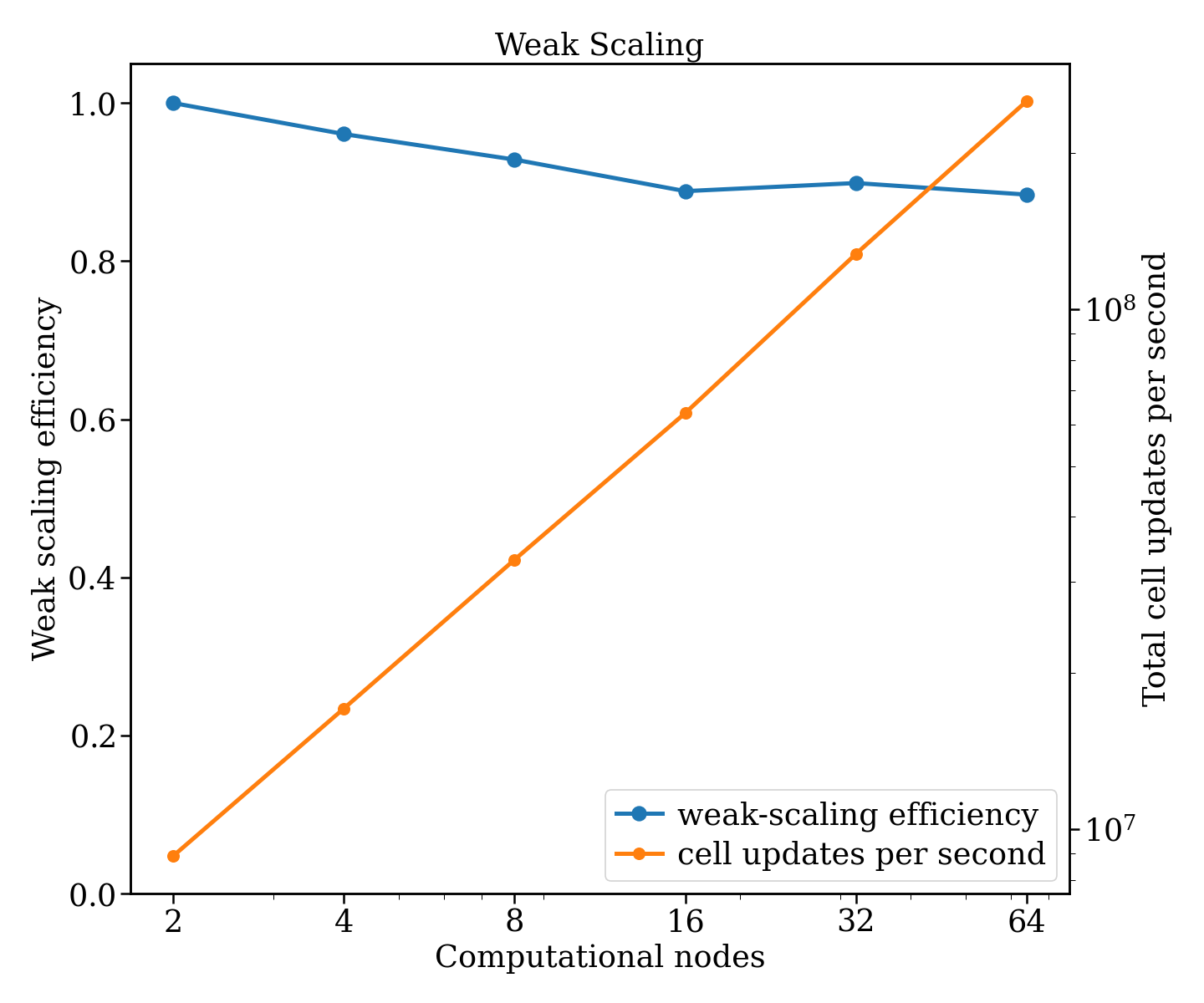}
  \caption{Measured weak-scaling efficiency and number of RK4 cell-updates per second at given node count on Perlmutter. Each node contains four Nvidia A100 GPUs. The measured weak-scaling efficiency remains above 80\% up to 64 nodes (256 GPUs).}
  \label{fig:weak-scaling}
\end{figure}

The weak-scaling test is done with the block size fixed to 32 and the active cell count per GPU fixed to 1,048,576. The number of blocks in the domain runs from 256 to 8192, while the number of computation nodes are correspondingly increased from 2 to 64. Each node contains four Nvidia A100 GPUs. Figure \ref{fig:weak-scaling} shows the measured weak-scaling efficiency and number of RK4 cell-updates per second at given node count. The weak-scaling efficiency remains above 85\% up to 64 nodes (256 GPUs). 

\section{Summary}
\label{sec:summary}

This paper presents {\sc Chana}, a block-structured adaptive-mesh code which evolves general-relativistic magnetohydrodynamics, multi-group M1 neutrino transport, and a dynamical spacetime in the Z4c formulation as one coupled system. Separate ghost widths and separate reconstruction/prolongation/restriction methods allow each physical module to use the stencil appropriate to its numerical method. The implementation is parallelized with MPI and Kokkos and uses flattened block-cell kernels, coalesced structure-of-arrays storage, kernel fusion, and device-side halo packing for efficient execution on GPUs.

A few design choices of the radiation scheme are worth emphasizing again: (a) The Lorentz-covariant Levermore closure is evaluated algebraically in the Eulerian frame, avoiding an additional nonlinear solve. The analytic expression of the Levermore closure allows one to compute its derivatives very easily, which unlocks the next design choice. (b) The stiff matter-radiation interactions are solved implicitly in each RK stage, and importantly, the Jacobian is derived using the chain rule instead of numerical differencing. This avoids the expensive repeated residual calculations, table readings, and closure evaluations during the Newton-Raphson solve. This choice is in particular helpful for GPU performance, because a large fraction of memory-bandwidth-limited scattered table readings are replaced by computation-heavy chain-rule derivative calculations. (c) The second-order method of \citet{radice2022} is used to compute spatial radiation fluxes because it's deliberately robust and inexpensive. Although the shadow and curved-beam tests show the corresponding numerical diffusion near sharp radiation features in optically-thin regions, such features are rarely found in astrophysical systems {\sc Chana} intends to model. The HLL fluxes computed from reconstructed face states are also implemented, which can reproduce sharper features when combined with higher-order reconstruction methods. However, the \citet{radice2022} fluxes are used by default because of their performance advantages.

The tests exercise the principal components of the code.  Linear MHD and gravitational waves converge at the expected spatial order. The Kelvin-Helmholtz instability and relativistic magnetized blast reproduce their expected structures, and the uniform- and adaptive-mesh blast solutions agree to the accuracy selected by the refinement criterion. The radiation module reproduces the analytic diffusion of a pulse moving at $0.5c$, which verifies both the diffusion-limit behavior of the radiation fluxes and the radiation-matter coupling. The radiation module also reproduces the expected Doppler and gravitational energy shifts, attenuation behind an absorbing cylinder, and the bending of a beam along null characteristics.  In addition, the binary-puncture calculation and neutron star collapse tests validate the code's capability to evolve dynamic spacetimes.  
Finally, the performance is measured for a representative three-species, 12-group CCSN GRRMHD configuration with Nvidia A100 GPUs on Perlmutter. When size-32 blocks are used, the performance rates are approximately $4.4\times10^5$ cell updates per GPU second with eSSPRK$(10,4)$ and $1.1\times10^6$ with RK4 time integrators. The per-step cost is roughly proportional to the number of RK stages per cell update. The weak-scaling efficiency remains above 85\% up to 256 GPUs (64 Perlmutter nodes).

Future development of the code includes (but are not limited to): improving the approximate implicit solver of radiation-matter coupling which provides the candidate solutions, overlapping ghost halo exchange more with kernel work on active cells, and improving GPU utilization of each individual kernel. There might be another factor of 1.5 to 2 speedup by changing the time integration scheme. \citet{Kuroda2026} evolves the Z4c and radiation-hydrodynamics sectors with different time integrators: Z4c with four Runge-Kutta stages, others with two. This means the per-step cost of the most expensive radiation sector is halved, compared to the standard four-stage RK4 method. However, it is unclear how this scheme may impact the robustness of the simulation when spacetime is evolving rapidly, and further tests are desired.

Code availability: {\sc Chana} is still being developed actively, the author will only provide the code upon collaboration or requests at this moment. After further testing and debugging and more performance optimizations, the code will become publicly available.

\section*{Acknowledgments}
The development and optimization of the code involves LLM tools including various versions/models of Codex, Claude Code, Gemini, GLM, and OpenCode. The author thanks the many open source projects (e.g., Athena++, AthenaK, Phoebus, WhiskyTHC, etc) and their contributors. The author (and likely the LLM tools) has learned a lot from such projects. The author would like to thank Daniel Kasen, Adam Burrows, and David Radice for the very helpful discussions. The author acknowledges support by the U.~S.\ Department of Energy under grant DE-SC0004658, support by the Gordon and Betty Moore Foundation through Grant GBMF5076, and support through a Simons Foundation grant (622817DK). The author acknowledges the Savio computational cluster resource provided by the Berkeley Research Computing program at the University of California, Berkeley; computational resources provided by the high-performance computer center at Princeton University, which is jointly supported by the Princeton Institute for Computational Science and Engineering (PICSciE) and the Princeton University Office of Information Technology; and allocation at the National Energy Research Scientific Computing Center (NERSC), which is supported by the Office of Science of the U.~S.\ Department of Energy under contract DE-AC03-76SF00098. The author acknowledges access to the Frontera cluster (under awards AST20020 and AST21003). This research is part of the Frontera computing project at the Texas Advanced Computing Center \citep{Stanzione2020}. Frontera is made possible by NSF award OAC-1818253. Additionally, a generous award of computer time was provided by the INCITE program, enabling this research to use resources of the Argonne Leadership Computing Facility, a DOE Office of Science User Facility supported under Contract DE-AC02-06CH11357.

\bibliographystyle{aasjournal}
\bibliography{References}

\label{lastpage}

\end{document}